\documentclass[10pt,a4paper,twocolumn]{article}
\usepackage[T1]{fontenc}
\usepackage[utf8]{inputenc}
\usepackage{tgtermes,tgheros}

\usepackage[a4paper,top=18mm,bottom=18mm,left=18mm,right=18mm,columnsep=6mm]{geometry}
\usepackage{amsmath,amssymb,bm,graphicx,booktabs,array,caption,subcaption,tikz,fancyhdr}
\usepackage[numbers,sort&compress]{natbib}
\usepackage[expansion=false]{microtype}
\usepackage{xurl}
\usepackage[hidelinks]{hyperref}
\graphicspath{{figures/}}
\makeatletter
\renewcommand{\maketitle}{%
  \begingroup\centering
  {\large\bfseries\@title\par}%
  \ifx\@author\@empty\else
    \vspace{2mm}{\normalsize\begin{tabular}[t]{c}\@author\end{tabular}\par}%
  \fi
  \ifx\@date\@empty\else\vspace{1mm}{\small\@date\par}\fi
  \vspace{3mm}\endgroup}
\makeatother
\renewenvironment{abstract}{\small\noindent\textbf{Abstract.}\ }{\par}

\usepackage{color}

\usetikzlibrary{arrows.meta,positioning}
\definecolor{cyclecolor}{rgb}{.90,.40,.02}
\definecolor{stagecolor}{rgb}{.78,.10,.16}
\definecolor{nomcolor}{rgb}{.06,.36,.70}
\definecolor{pidcolor}{rgb}{.35,.39,.45}
\definecolor{rlcolor}{rgb}{.02,.48,.40}
\definecolor{fixedgaincolor}{rgb}{.40,.22,.65}
\newcommand{\keyline}[2]{\textcolor{#1}{\rule{5mm}{1pt}}\,#2\quad}
\newcommand{\curvekey}[4]{\tikz[baseline=-.6ex]{\draw[color=#1,#2,line width=.8pt] (0,0)--(.55,0);\node[fill=white,inner sep=.5pt,font=\tiny] at (.275,0) {#3};}\,#4\quad}
\newcommand{\basiclegend}{\par\smallskip{\small\curvekey{pidcolor}{dashed}{$\square$}{PID}\curvekey{nomcolor}{dash dot}{$\diamond$}{Nominal}\curvekey{cyclecolor}{dashed}{$\triangle$}{Supervised cyclewise}\curvekey{stagecolor}{solid}{$\circ$}{Supervised stagewise}}}
\newcommand{\cfdlegend}{\par\smallskip{\small\curvekey{pidcolor}{dashed}{$\square$}{PID}\curvekey{nomcolor}{dash dot}{$\diamond$}{Zero residual}\curvekey{cyclecolor}{dashed}{$\triangle$}{Cyclewise}\curvekey{stagecolor}{solid}{$\circ$}{Stagewise}\curvekey{rlcolor}{dotted}{$\times$}{Internal stagewise RL}}}

\definecolor{predL}{rgb}{.05,.40,.72}
\definecolor{predC}{rgb}{.08,.56,.40}
\definecolor{predF}{rgb}{.86,.40,.04}
\newcommand{\predictionlegend}{\par\smallskip{\centering\small
\tikz[baseline=-.5ex]{\fill[predL] (0,0) rectangle (.18,.14);}\,L\quad
\tikz[baseline=-.5ex]{\fill[predC] (0,0) rectangle (.18,.14);}\,C\quad
\tikz[baseline=-.5ex]{\fill[predF] (0,0) rectangle (.18,.14);}\,F8\qquad
\tikz[baseline=-.5ex]{\draw[line width=.8pt] (.09,.07) circle (.065);}\,Pass: $\mathrm{RMSE}_{g_\phi}\leq\mathrm{RMSE}_0$\quad
\tikz[baseline=-.5ex]{\draw[line width=.8pt] (.02,0)--(.16,.14) (.02,.14)--(.16,0);}\,Fail: $\mathrm{RMSE}_{g_\phi}>\mathrm{RMSE}_0$\par}}
\newcommand{\rllegend}{\par\smallskip{\small\curvekey{cyclecolor}{solid}{$\circ$}{Cyclewise baseline}\curvekey{stagecolor}{solid}{$\circ$}{Stagewise baseline}\curvekey{rlcolor}{dashed}{$\square$}{Matched RL}}}

\title{Pre-Solve Horizon Sampling of Frozen Acceleration Residuals for Quadrotor NMPC under External Forces}
\author{Sun Shenghan\\
Department of Mechanical and Aerospace Engineering\\
School of Engineering\\
The Hong Kong University of Science and Technology\\
Hong Kong SAR, China}
\date{}

\begin{document}
\twocolumn[{\maketitle
\begin{minipage}{\textwidth}
\begin{abstract}
We investigate how a frozen supervised acceleration-residual model should be sampled within nonlinear model predictive control (NMPC) of a quadrotor subject to external forces. Cyclewise deployment holds the current-state residual across the prediction horizon, whereas stagewise deployment evaluates the same network at shifted warm-start nodes before each optimization solve. The correction sequence is fixed during optimization and the network is not updated online. In 220 simulated pulse-load conditions, cyclewise and stagewise deployment reduce mean integral absolute position error (IAE) relative to nominal NMPC by 4.38\% and 7.77\%, respectively. However, both worsen all 55 hover conditions, and most of the aggregate improvement arises after pulse removal rather than during loading. In 96 prescribed random-force blocks, stagewise deployment reduces mean IAE by 11.99\%; the evaluated finite-budget reinforcement-learning add-ons do not improve their matched supervised means in either force bank. A separate causal innovation compensator reduces mean IAE by 27.47\% relative to stagewise deployment on 32 confirmation conditions, at the cost of a 10.19\% increase in angular-rate root-mean-square error. In one synthetic computational-fluid-dynamics inflow field, stagewise deployment reduces four-task mean IAE by 2.01\% against a same-graph zero-residual control. These simulation results quantify deployment-dependent tracking gains and trade-offs, without establishing general wind rejection or hard-real-time guarantees.
\end{abstract}

\end{minipage}\vspace{3mm}}]

\section{Introduction}\label{intro}

Quadrotor trajectory tracking requires a controller to reconcile a predictive model with aerodynamic loads, actuator dynamics and external disturbances. A nominal rigid-body model can provide an efficient optimization backbone, yet leave motion-dependent accelerations unmodeled. Drag-aware reference construction and data-driven predictive models have improved tracking in their evaluated regimes \cite{r1,r3}. These advances motivate a compact learned correction to the model. The practical question is how that correction remains useful after horizon prediction, optimization and feedback interact in the executed trajectory.

Nonlinear model predictive control (NMPC) couples future reference information to actuator-constrained prediction. Its benefits depend on both the model and how model information enters an affordable real-time solve \cite{r4,r5}. A larger neural model is not automatically a better deployed predictor: accuracy, derivative evaluation and runtime interact \cite{r7}. A frozen acceleration residual offers a controlled alternative. It preserves the nominal dynamics and uses a supervised network only for the missing linear and angular accelerations. The network can be trained separately from controller deployment, allowing a comparison in which the data and weights remain unchanged.

Even with fixed weights, residual deployment is not unique. One can evaluate the network at the measured state once per control cycle and hold its output across the prediction window. Alternatively, one can sample shifted warm-start nodes before the solve and supply a different correction at each node. The first rule reduces preparation; the second represents predicted variation in velocity and rotor state but relies on an approximate forecast. Neither guarantees better execution. Data-Driven MPC learns Gaussian-process acceleration corrections \cite{r3}; Real-Time Neural MPC supplies learned-model values and derivatives to its local optimization approximation \cite{r7}. Here the comparison isolates two ways to supply values from one frozen network, with no learned-model derivatives entering the solve. The contribution is NOT a new neural architecture, reinforcement-learning algorithm, or disturbance estimator. It is a controlled comparison of two pre-solve sampling rules for the same frozen acceleration-residual model. Both use identical training data and weights, and both hold the sampled correction sequence fixed during each SQP-RTI solve. This isolates a practical deployment decision from model fitting, while leaving warm-start forecast quality and solver sensitivity as possible mediators.

Prediction fit is also an incomplete selection criterion. Supervision minimizes a model loss, while the controller is judged by accumulated tracking error, transient peaks and terminal recovery. Control-oriented identification explicitly distinguishes prediction objectives from closed-loop performance \cite{r39}. A small componentwise residual error can matter differently in hover and a moving trajectory; an improved cross-task average can conceal a degraded static condition. A controller can also reduce integrated error by recovering faster after a force is removed without reducing the largest disturbance excursion. Separate model assessment, taskwise comparison and temporal decomposition are needed to distinguish these possibilities.

Learned policies provide a further benchmark for the residual model. Residual reinforcement learning (RL) augments an existing controller rather than replacing it \cite{r15}. An external policy can modify the optimized motor command; an internal policy can alter the dynamics correction presented to NMPC. Those interfaces change different parts of the executed model--control chain and must be compared against the supervised method each actually modifies. Soft actor-critic (SAC) optimizes an entropy-regularized discounted return \cite{r28,r35}, whereas the tracking score here is an undiscounted position-error integral. A rising return or a more coherent interface alone cannot establish the additional closed-loop benefit of a deployed policy.

Causal innovation compensation supplies a useful non-RL comparison. The discrepancy between the preceding predicted transition and the measured state is information unavailable to a frozen still-air model. Filtering and feeding back this discrepancy follows a disturbance-observer principle \cite{r41}, while the resulting innovation can contain intrinsic model and integration error as well as applied loading. Its position-tracking benefit must be considered alongside angular-motion costs. This comparison prevents a learned method from being judged only against an uncompensated reference, and prevents a gain in one metric from being called overall superiority.

This paper answers three questions under controlled numerical conditions:
\begin{enumerate}
\item[Q1.] How do frozen residual deployment and added external or internal policies affect tracking relative to their matched baselines?
\item[Q2.] What position and angular-motion trade-off follows from causal innovation compensation?
\item[Q3.] How do these observations change under time-varying forces and CFD rotor inflow, and what computation is required?
\end{enumerate}
The 220-condition pulse grid, separately generated 32-condition confirmation, random-force bank and fixed CFD field remain distinct comparisons. Archived PID trajectories enter through an established plant and case bridge; the policy identity is fixed in the experimental design; the other trained seeds remain in the engineering archive. The CFD velocity modifies only the common plant: no predictor or actor receives the field, and no weights are retrained. Thus it tests limited environment transfer rather than a new CFD-informed controller. The retained structural interventions do not replace the original benchmarks.

The three contributions are:
\begin{enumerate}
\item \textbf{Two deployment schemes for one frozen supervised residual model}: cyclewise deployment evaluates $g_\phi$ once at the current body-frame velocity, angular velocity and actual rotor-speed state, then holds its correction over the NMPC horizon. Stagewise deployment evaluates the same network at shifted warm-start nodes to represent predicted state variation. Both sequences are fixed before optimization, without network derivatives in the solve or weight updates during control. This comparison tests horizon sampling at unchanged training data and weights.
\item A \textbf{closed-loop evaluation} against nominal control, policy corrections and causal compensation, reporting task-dependent tracking, angular-motion costs and computation separately.
\item A \textbf{CFD-environment validation}: five controllers are compared on four reference trajectories in one prescribed turbulent velocity field. The field enters only the common plant, with no CFD inputs to predictors or actors and no retraining. The comparison tests transfer of the frozen supervised schemes while reporting position and angular-motion outcomes separately (Figure~\ref{fig:cfd}).
\end{enumerate}

\section{Related Work}\label{related}

\subsection{Aerodynamics and learned prediction}

Drag-aware flatness \cite{r1}, adaptive NMPC \cite{r6}, data-driven MPC \cite{r3}, and neural MPC \cite{r7} address different parts of the model--control chain. Onboard learning-based MPC provides an implementation precedent \cite{r8}; NeuroBEM combines physics and learned aerodynamic residuals \cite{r17}. These results motivate residual augmentation, but do not supply the outcome of our frozen comparison. Control-oriented identification distinguishes prediction fit from performance under feedback \cite{r39}; both must be measured here.

\subsection{Residual policy and fixed compensation}

Residual RL augments a baseline command \cite{r15}; other MPC--RL combinations constrain learning or tune controller parameters \cite{r20,r21}. SAC uses maximum-entropy off-policy learning \cite{r28,r35}, with a versioned implementation in Stable-Baselines3 \cite{r36,r38}. Its benchmark success does not guarantee an increment over a strong local baseline. Few-run evaluation warrants separate seed reporting \cite{r37,r40}. Disturbance-observer control motivates a causal compensation channel \cite{r41}, but an acceleration innovation is not automatically an identified external force.

Demonstration-based initialization \cite{r42} and frozen-policy refinement \cite{r43} motivate preserving an existing controller during learning. The residual-value-learning preprint \cite{r44} proposes specific cold-start and scale-mismatch explanations in other robotic tasks. We use it to formulate a possible explanation, not to diagnose our critic or claim implementation of that complete algorithm. The longer methodological comparison is retained in Supplementary Section~\ref{sec:extended_related}.

\section{Methods}\label{sec:methods}

\begin{figure*}[!t]
\centering\begin{tikzpicture}[x=1cm,y=1cm,>=Stealth,
box/.style={draw=gray,rounded corners=1mm,minimum height=9mm,text width=3.1cm,align=center,font=\small},
warm/.style={box,fill=stagecolor!9},cold/.style={box,fill=nomcolor!6}]
\node[cold] (data) at (1.8,2) {Still-air data\\$(\boldsymbol\chi,\mathbf y)$};
\node[warm] (fit) at (7.4,2) {Fit residual model\\$g_\phi(\boldsymbol\chi)$};
\node[warm] (freeze) at (13,2) {Freeze model\\$\phi=\mathrm{constant}$};
\node[cold,text width=2.8cm] (state) at (1.7,0) {State $\mathbf x_k$, reference\\$\mathbf x_{\rm ref}$, warm start $\tilde{\mathbf x}_{i|k}$};
\node[warm,text width=3.2cm] (eval) at (5.8,0) {Evaluate $g_\phi$\\$\mathbf g_{i|k}$ (before solve)};
\node[warm,text width=2.6cm] (solve) at (9.9,0) {Supervised NMPC\\$f_c=f_0+B\mathbf g_{i|k}$};
\node[cold,text width=2.4cm] (plant) at (13.7,0) {First motor command\\$\mathbf u_{0|k}^{\star}$; plant};
\draw[->] (data)--(fit);\draw[->] (fit)--(freeze);
\draw[->] (freeze.south)--(13,1)--(eval.north);
\draw[->] (state)--(eval);\draw[->] (eval)--(solve);\draw[->] (solve)--(plant);
\draw[->] (plant.south)--(13.7,-1)--(1.7,-1)--(state.south);
\node[font=\small] at (7.5,-1.35) {Measured-state feedback $\mathbf x_{k+1}$};
\end{tikzpicture}
\caption{The proposed supervised residual NMPC separates still-air model fitting from frozen online deployment. The main stagewise rule samples the same network at shifted warm-start prediction nodes before optimization; the cyclewise comparator repeats one current-state output. Neither rule receives the imposed force or changes network weights during control.}\label{fig:fig01_framework}
\end{figure*}

Figure~\ref{fig:fig01_framework} separates offline fitting from online control: feedback changes the sampled state, not the network weights. Both deployments use the same $g_\phi$ before the NMPC solve; only the horizon sampling rule changes. The diagram therefore defines a deployment comparison, rather than two independently learned models.
\subsection{State, loads, and nominal prediction}

Position $\mathbf p$ and velocity $\mathbf v$ use the east--north--up world frame (W); angular velocity $\boldsymbol\omega$ uses front--left--up body axes (B). The unit quaternion $\mathbf q=\operatorname{col}(\mathbf q_v,q_w)$ uses xyzw order. $R(\mathbf q)$ maps B to W. Actual rotor speeds $\boldsymbol\Omega$ differ from commanded speeds $\mathbf u$. For column vectors, $\operatorname{col}(\mathbf a,\mathbf b)=[\mathbf a^T,\mathbf b^T]^T$ denotes vertical concatenation. Superscript $T$ denotes transpose.

\begin{equation}\label{eq:state}\mathbf x=\operatorname{col}(\mathbf p,\mathbf v,\mathbf q,\boldsymbol\omega,\boldsymbol\Omega)\in\mathbb R^{17},\quad\mathbf u\in\mathbb R^4.\end{equation}

\begin{equation}\label{eq:translation}\dot{\mathbf p}=\mathbf v,\quad m\dot{\mathbf v}=-mg\mathbf e_3+R(\mathbf q)\mathbf F^B+\mathbf F_{\rm ext}^{W}.\end{equation}

\begin{equation}\label{eq:rotation}I\dot{\boldsymbol\omega}=\mathbf M^B-\boldsymbol\omega\times(I\boldsymbol\omega)+\mathbf M_{\rm ext}^{B}.\end{equation}

\begin{equation}\label{eq:quaternion}\begin{aligned}\dot{\mathbf q}_v&=\tfrac12(q_w\boldsymbol\omega+\mathbf q_v\times\boldsymbol\omega),\\\dot q_w&=-\tfrac12\mathbf q_v^T\boldsymbol\omega,\quad\|\mathbf q\|_2=1.\end{aligned}\end{equation}

\begin{equation}\label{eq:motor}\dot\Omega_j=(u_j-\Omega_j)/\tau_j,\quad j=1,\ldots,4.\end{equation}

In Eqs.~\eqref{eq:state}--\eqref{eq:motor}, $\mathbf p,\mathbf v,\boldsymbol\omega\in\mathbb R^3$, $\mathbf q\in\mathbb R^4$, $\boldsymbol\Omega\in\mathbb R^4$, and $R,I\in\mathbb R^{3\times3}$. Here $m$, $g$, $I$, and $\mathbf e_3=(0,0,1)^T$ denote mass, gravity, positive-definite body inertia, and the vertical unit vector. Rotor loads $\mathbf F^B,\mathbf M^B$ exclude injected environmental loads. The plant includes axial thrust attenuation, transverse rotor drag, rolling moments, full inertial coupling, and asymmetric motor lag. The nominal model $f_0$ retains quadratic thrust and reaction torque, rigid-body coupling, and motor lag, but omits those aerodynamic differences. Thus mismatch exists without injected force. Actuator dynamics and asymmetric limits matter in multirotor NMPC \cite{r30}. Here the adopted frozen vehicle model uses the acceleration constant for $\tau_j$ when $u_j>\Omega_j$ and the deceleration constant otherwise; the model-specific values are listed in Supplementary Section~\ref{sec:settings}.

\subsection{Baseline controllers}

PID (E0) uses current position and velocity references, bounded integral state $\boldsymbol\xi$, and diagonal gain matrices. Gravity compensation and a desired yaw construct the desired attitude; attitude feedback and allocation produce four motor commands \cite{r22,r23,r31}. The candidate gain derivation and anti-windup rule are given in Supplementary Section~\ref{sec:pid_details}; Table~\ref{tab:pid_gains} lists the gains.

\begin{equation}\label{eq:pid}\mathbf a_{\rm PID}=K_p(\mathbf p_{\rm ref}-\mathbf p)+K_i\boldsymbol\xi+K_v(\mathbf v_{\rm ref}-\mathbf v).\end{equation}

\subsection{NMPC cost and corrected dynamics}

Nominal NMPC (E1) uses future references over $N$ intervals of length $\Delta$. Index $k$ counts control cycles; $i|k$ denotes prediction node $i$ formed at cycle $k$, and $k-1$ denotes the preceding cycle. A star denotes the command returned by optimization, with $\mathbf u_k^{\rm sup}=\mathbf u_{0|k}^{\star}$ for the supervised solve. Bold symbols denote vectors or matrices; superscripts W and B indicate world and body coordinates. The control-output map $\eta:\mathbb R^4\to\mathbb R^4$ returns commanded total thrust $T_c$ (N) and body moments $M_x,M_y,M_z$ (N m). The auxiliary four-vector $\mathbf h$ stores the preceding commanded thrust and moment, $\eta(\mathbf u)=\operatorname{col}(T_c,M_x,M_y,M_z)$. The augmented optimization state has 21 entries. At cycle $k$, $\mathbf h_{0|k}=\eta(\mathbf u_{k-1}^{\rm exec})$ and $\mathbf h_{i+1|k}=\eta(\mathbf u_{i|k})$. The cost is

\begin{equation}\label{eq:cost}\begin{aligned}J_k&=\sum_{i=0}^{N-1}\ell_{i|k}+\ell_{N|k},\\\ell&=\|\mathbf e_x\|_2^2+w_u\|D_u^{-1}(\eta(\mathbf u)-\eta_{\rm ref})\|_2^2\\&\quad+w_{\Delta u}\|D_u^{-1}(\eta(\mathbf u)-\mathbf h)\|_2^2,\\\ell_{N|k}&=w_N\|\mathbf e_{x,N|k}\|_2^2.\end{aligned}\end{equation}

In Eq.~\eqref{eq:cost}, the 12-vector $\mathbf e_x$ concatenates scaled position, velocity, quaternion-vector attitude, and angular-rate errors. Its attitude term is $2\operatorname{vec}(\mathbf q_{\rm ref}^{-1}\otimes\mathbf q)$; $\operatorname{vec}$ extracts the three-vector part and $\otimes$ denotes quaternion multiplication. $D_u$ is the positive diagonal thrust--moment scale. The optimizer minimizes $J_k$ subject to the discrete dynamics and commanded/actual motor bounds; only its first command is executed \cite{r4,r5}. Position is not a hard optimization constraint.

\subsection{Supervised residual and its two evaluation rules}

\begin{equation}\label{eq:features}\boldsymbol\chi=\operatorname{col}(R(\mathbf q)^T\mathbf v,\boldsymbol\omega,\boldsymbol\Omega)\in\mathbb R^{10}.\end{equation}

\begin{equation}\label{eq:labels}\mathbf y=\operatorname{col}(R(\mathbf q)^T\delta\dot{\mathbf v},\delta\dot{\boldsymbol\omega})\in\mathbb R^6.\end{equation}

In Eqs.~\eqref{eq:features}--\eqref{eq:labels}, the label differences $\delta\dot{\mathbf v},\delta\dot{\boldsymbol\omega}$ are plant-minus-nominal accelerations at the same state and input. They are instantaneous residuals, not position errors or one-step transition errors. Still-air collection has no injected force or torque. Body-expressed world velocity $R^T\mathbf v$ is not wind-relative velocity. The rotor block is the actual-speed state $\boldsymbol\Omega$, represented online by the propagated actuator estimate; the command $\mathbf u$ affects motor dynamics but is not an additional network feature. A frozen 10--32--6 tanh network and its training-set normalization yield physical outputs $g_{\rm raw}$. Componentwise normalization balances the differently dimensioned inputs and targets during fitting. Its statistics are fitted only to the training rows. Fitting minimizes the mean squared error over all six standardized targets, with no additional component weights; Eq.~\eqref{eq:fit_loss} gives the exact objective. Inverse scaling restores physical acceleration units:
\begin{equation}\label{eq:normalization}
g_{\rm raw}(\boldsymbol\chi)=\boldsymbol\mu_y+\boldsymbol\sigma_y\odot G_\phi\big((\boldsymbol\chi-\boldsymbol\mu_\chi)\oslash\boldsymbol\sigma_\chi\big).
\end{equation}
Here $G_\phi:\mathbb R^{10}\to\mathbb R^6$ is the normalized-output network; $\boldsymbol\mu_\chi,\boldsymbol\sigma_\chi\in\mathbb R^{10}$ and $\boldsymbol\mu_y,\boldsymbol\sigma_y\in\mathbb R^6$ are componentwise training means and positive normalization scales. $\odot$ and $\oslash$ denote componentwise multiplication and division. The first three physical outputs are m/s$^2$, the last three rad/s$^2$. Mirror averaging uses

\begin{equation}\label{eq:mirror}g_\phi(\boldsymbol\chi)=\tfrac12[g_{\rm raw}(\boldsymbol\chi)+Dg_{\rm raw}(\mathcal T\boldsymbol\chi)].\end{equation}

$\phi$ contains frozen weights and biases. $\mathcal T$ mirrors velocity by $Q=\operatorname{diag}(1,-1,1)$, angular velocity by $-Q$, and motor order to $(3,4,1,2)$; $D=\operatorname{diag}(1,-1,1,-1,1,-1)$ acts on the six outputs. This paired-rotor still-air symmetry is not a symmetry assumption on arbitrary wind. A body-frame correction $\mathbf d=\operatorname{col}(\mathbf d_v,\mathbf d_\omega)$ enters continuous dynamics before integration:

\begin{equation}\label{eq:corrected}\begin{aligned}f_c(\mathbf x,\mathbf u;\mathbf d)&=f_0(\mathbf x,\mathbf u)+B(\mathbf q)\mathbf d,\\B(\mathbf q)\mathbf d&=\operatorname{col}(\mathbf0_3,R\mathbf d_v,\mathbf0_4,\mathbf d_\omega,\mathbf0_4),\\F_\Delta&=\operatorname{RK4}_{\Delta,8}[f_c].\end{aligned}\end{equation}

For cyclewise sampling, the network is evaluated once at the measured state and its output is repeated across all N prediction stages. For stagewise sampling, N outputs are evaluated at the shifted warm-start states before each SQP-RTI solve. These values are numerical parameters to the optimizer; the network is not re-evaluated at new SQP trial states and contributes no learned-model derivatives to the optimization linearization. The terminal node copies the final stage.

The two rules address a computational choice inherent in residual deployment. Holding a current-state correction is inexpensive but ignores changes in velocity, angular rate and rotor speed along a moving prediction. Sampling the warm start represents those changes using the same model and training data, at the cost of more inference and sensitivity to a forecast that is not the final optimized trajectory. This motivates the comparison without assuming that more samples guarantee better control. Let $\tilde{\mathbf x}_{i|k}$ denote the shifted previous solution, initialized from the measured state at $i=0$, and let $\chi(\mathbf x)$ be Eq.~\eqref{eq:features}. Then
\begin{equation}\label{eq:rules}
\mathbf g_{i|k}=\begin{cases}
g_\phi(\chi(\mathbf x_k)),&\text{cyclewise},\\
g_\phi(\chi(\tilde{\mathbf x}_{i|k})),&\text{stagewise},
\end{cases}\quad 0\le i<N.
\end{equation}
At the first cycle there is no preceding solution: the supervised controller rolls out the previous command with the current residual held fixed, then samples the resulting nodes for stagewise deployment. Subsequently it shifts the preceding successful state and command sequence, repeats its final node, and overwrites the first state with the current estimate. Supplementary Section~\ref{sec:prediction_details} gives the initialization and collection controller. The main deployment is stagewise; cyclewise is a controlled evaluation of the same frozen model. Its selection is assessed through closed-loop error and computation, rather than justified by prediction fit alone.

\subsection{Matched policy benchmarks and additional causal compensation}

\begin{equation}\label{eq:external}\mathbf u_k^{\rm exec}=\operatorname{sat}(\mathbf u_k^{\rm sup}+0.02\boldsymbol\Omega_{\rm hover}\odot\mathbf a_k).\end{equation}

\begin{equation}\label{eq:internal}\mathbf d_{i|k}=\mathbf g_{i|k}+P_i\mathbf b_k,\quad\mathbf b_k=\mathbf s_b\odot\mathbf a_k.\end{equation}

\begin{equation}\label{eq:action}\mathbf a_k=\operatorname{clip}(\pi_\theta(\mathbf o_k),-1,1).\end{equation}

External RL (E3) has four normalized motor actions. Internal RL (E4) has six body-acceleration actions. In Eq.~\eqref{eq:internal}, $\mathbf s_b,\mathbf b_k\in\mathbb R^6$ contain linear (m/s$^2$) and angular (rad/s$^2$) blocks, and $P_i\in\mathbb R^{6\times6}$; $\odot$ is componentwise multiplication, and $\operatorname{sat}$ enforces motor bounds. In Eq.~\eqref{eq:external}, $\boldsymbol\Omega_{\rm hover}\in\mathbb R^4$ is the equilibrium motor-speed vector. Equation~\eqref{eq:action} defines the deterministic normalized action of actor $\pi_\theta$, with parameter vector $\theta$ and observation $\mathbf o_k$. The 24-/26-component observations comprise unscaled position, velocity, and angular-rate differences, a sign-canonical four-component attitude-error quaternion, preceding executed command, preceding 4-/6-action history, and reference velocity. Unknown force, case identity, absolute time, and future realized states are excluded. Only executed commands and action history are normalized in these original observations.  The exact concatenation is in Supplementary Section~\ref{sec:rl_details}.

The original internal policies use $P_i=I_6$: a cycle's action is held over the prediction horizon. The external policies modify the cyclewise supervised controller; the internal policies are evaluated with each supervised rule at fixed actor weights. These are benchmarks for an additional policy increment, rather than components required by the proposed supervised method. Fixed-policy structural tests and residual removal/retraining are defined in Supplementary Section~\ref{sec:structure}.

\begin{equation}\label{eq:observer}\begin{aligned}\boldsymbol\nu_k^W&=(\mathbf v_k^W-\hat{\mathbf v}_{k|k-1}^W)/\Delta,\\\hat{\mathbf d}_k^W&=(1-\beta)\hat{\mathbf d}_{k-1}^W+\beta\boldsymbol\nu_k^W,\\\mathbf b_{v,k}^{B}&=\operatorname{clip}(\kappa R^T\hat{\mathbf d}_k^W,-\mathbf s_v,\mathbf s_v),\\\mathbf b_{i|k}&=\rho^i\operatorname{col}(\mathbf b_{v,k}^{B},\mathbf0_3).\end{aligned}\end{equation}

In Eq.~\eqref{eq:observer}, $\Delta$ is the control interval, $\beta$ the filter coefficient, $\kappa$ the gain, and $\rho$ the horizon attenuation. The fixed compensator predicts the previous executed transition under $f_0+g_\phi$ without added compensation. Here $\hat{\mathbf v}_{k|k-1}^W$ is the velocity predicted for cycle $k$ from cycle $k-1$; a hat marks a prediction or estimate, while the tilde in Eq.~\eqref{eq:rules} marks a warm-start state. Its innovation uses only information available at cycle $k$, has acceleration units, and may include model/discretization error as well as external loading. It is not identified force. The stagewise supervised term is retained: the fixed model correction is $\mathbf d_{i|k}=\mathbf g_{i|k}+\mathbf b_{i|k}$. Thus this comparator adds online innovation compensation to the supervised model; it does not replace supervision or use a learned actor. The compensation changes the predicted acceleration; the NMPC optimizer then maps the corrected prediction to motor-speed commands under the same dynamics and actuator constraints. $\hat{\mathbf v}_{k|k-1}^W$, $\boldsymbol\nu_k^W$, and $\hat{\mathbf d}_k^W$ are three-vectors; $\mathbf s_v$ is a three-vector of acceleration limits. The innovation is zero until a preceding executed transition exists, and its filter is reset between cases. Disturbance-observer control motivates the causal feedback of model discrepancy \cite{r41}; that principle does not establish that these particular gains are optimal. Figure~\ref{fig:interfaces} shows the resulting distinction: external RL changes the command after optimization, whereas internal RL and fixed compensation alter the model before optimization. In the latter two paths, NMPC computes the motor commands from the corrected prediction. This timing distinction defines the comparison; it does not by itself predict which interface will track better.

\subsection{CFD velocity as a plant-only input}\label{sec:cfd_method}
For supplementary environment validation, let $\mathbf w^W(\mathbf r,t)\in\mathbb R^3$ be the exported world-frame velocity (m/s) and $\mathbf r_j^B\in\mathbb R^3$ the body-frame arm vector (m) of rotor $j\in\{1,\ldots,4\}$. With the body-to-world rotation $R$, its location and relative air velocity are
\begin{equation}\label{eq:cfd_mapping}\begin{aligned}
\mathbf r_j^W&=\mathbf p^W+R\mathbf r_j^B,\\
\mathbf v_{{\rm air},j}^B&=R^T[\mathbf v^W-\mathbf w^W(\mathbf r_j^W,t)]
+\boldsymbol\omega^B\times\mathbf r_j^B.
\end{aligned}\end{equation}
Periodic trilinear spatial interpolation and linear interpolation between saved times supply the velocity at each rotor's current position. The existing source-frame rotor model uses this relative velocity and rotor speed for axial thrust attenuation, transverse drag and rolling moments; lever arms and yaw moments complete the plant wrench \cite{r29}. The environment is queried every 5 ms, with each query held for its plant segment. The predictor receives no CFD feature, the supervised weights remain frozen, and the original added center-of-mass pulse is zero. This changes the evaluation environment rather than the learned controller.

\subsection{Implementation and evidence boundaries}

CasADi, acados real-time iteration sequential quadratic programming (SQP-RTI), and the partially condensed High-Performance Interior-Point Method (HPIPM) implement the solver \cite{r24,r25,r26,r27,r32,r33}. Gauss--Newton is a cost-Hessian setting. Original SAC uses versioned Stable-Baselines3. Stochastic training and deterministic evaluation remain distinct \cite{r28,r35,r36,r38}. Numerical values, training budgets, reward definitions, and support boundaries are reported with the experimental design and supplementary settings. A successful solver return and rising training return are not certificates of tracking improvement.

\begin{figure*}[!t]
\centering
\begin{tikzpicture}[x=1cm,y=1cm,>=Stealth,
b/.style={draw=gray,rounded corners=1mm,text width=3.05cm,minimum height=11mm,align=center,font=\small},
c/.style={b,fill=nomcolor!6},w/.style={b,fill=stagecolor!8},a/.style={b,fill=rlcolor!6}]
\node[anchor=west,font=\small\bfseries] at (0,5.05) {External RL: motor correction after the NMPC solve};
\node[w] (es) at (1.75,3.65) {$g_\phi$ + NMPC\\solved command $\mathbf u_k^{\rm sup}$};
\node[a] (ea) at (5.8,3.65) {Actor $\pi_\theta(\mathbf o_k)$\\4 actions $\mathbf a_k$};
\node[c] (ec) at (9.85,3.65) {Add motor correction\\$0.02\boldsymbol\Omega_{\rm hover}\odot\mathbf a_k$};
\node[c,text width=2.5cm] (ep) at (13.65,3.65) {Motor bounds\\$\mathbf u_k^{\rm exec}$ $\to$ plant};
\draw[->] (ea)--(ec);\draw[->] (ec)--(ep);
\node[anchor=west,font=\footnotesize] at (0,2.70) {$\mathbf o_k$: current state/reference, previous executed command and action history. The solved command bypasses the actor.};
\draw[->,line width=.7pt] (es.north) -- (1.75,4.57) -- (9.85,4.57) -- (ec.north);

\node[anchor=west,font=\small\bfseries] at (0,2.2) {Internal RL: acceleration correction before the NMPC solve};
\node[a] (ia) at (1.75,1.3) {Actor $\pi_\theta(\mathbf o_k)$\\$\mathbf b_k=\mathbf s_b\odot\mathbf a_k\in\mathbb R^6$};
\node[w] (ig) at (5.8,1.3) {Frozen residual $g_\phi$\\$\mathbf d_{i|k}=\mathbf g_{i|k}+P_i\mathbf b_k$};
\node[w] (is) at (9.85,1.3) {NMPC\\$f_c=f_0+B\mathbf d_{i|k}$};
\node[c,text width=2.5cm] (ip) at (13.65,1.3) {First motor command\\$\mathbf u_{0|k}^{\star}$ $\to$ plant};
\draw[->] (ia)--(ig);\draw[->] (ig)--(is);\draw[->] (is)--(ip);
\node[anchor=west,font=\small\bfseries] at (0,.3) {Fixed compensation: causal prediction innovation, no actor};
\node[c] (fi) at (1.75,-.6) {Last executed transition\\$\boldsymbol\nu_k^W=(\mathbf v_k^W-\hat{\mathbf v}_{k|k-1}^W)/\Delta$};
\node[c] (ff) at (5.8,-.6) {Filter, gain, rotate, limit\\$\mathbf b_{i|k}=\rho^i\operatorname{col}(\mathbf b_{v,k}^B,\mathbf0_3)$};
\node[w] (fs) at (9.85,-.6) {$g_\phi$ + NMPC\\$\mathbf d_{i|k}=\mathbf g_{i|k}+\mathbf b_{i|k}$};
\node[c,text width=2.5cm] (fp) at (13.65,-.6) {First motor command\\$\mathbf u_{0|k}^{\star}$ $\to$ plant};
\draw[->] (fi)--(ff);\draw[->] (ff)--(fs);\draw[->] (fs)--(fp);
\end{tikzpicture}
\caption{Comparator interfaces preserve the NMPC backbone. External RL uses current state/reference and command/action history to modify the solved motor command. Internal RL inserts a six-dimensional acceleration correction before solving. Fixed compensation uses the last executed transition and the current velocity measurement; its prediction excludes the added compensation. The plant force is an evaluation input, never an observed policy or residual-model feature.}\label{fig:interfaces}
\end{figure*}

\section{Experimental Results}\label{sec:results}\subsection{Experimental design and settings}
All current comparisons use the assembled Iris plant (1.55 kg, $g=9.81$ m/s$^2$), 50 Hz control, 200 Hz state output, and a 20-interval/0.4 s prediction window. Plant RK45 and prediction RK4 have different step rules. Physical parameters, tolerances, costs, and learning configurations are fixed across each matched comparison; Supplementary Section~\ref{sec:settings} gives their exact values and reasons. Ideal simulated-state feedback and synchronous execution bound the conclusions.

The fixed grid contains hover H, line L, circle C, and figure-eight F8, with 55 cases each: one zero-force case plus six directions at nine positive intensities. The force is a world-frame center-of-mass pulse without added torque, not a physical wind field:

\begin{equation}\label{eq:force}\mathbf F_{\rm ext}^{W}(t)=\lambda mg\mathbf n_F\,\mathbf1_{[2,3)}(t).\end{equation}

$\lambda$ is force/weight and $\mathbf n_F$ a unit direction. The directions are $\pm X,\pm Y,(1,\pm1,0)/\sqrt2$; the intensities are 0.02, 0.04, 0.06, 0.08, 0.10, 0.12, 0.16, 0.20, and 0.24. This finite direction bank includes axis-aligned and two oblique loads; it is not a complete angular sweep, and no omitted direction is declared equivalent by symmetry. The 8 s window includes all 1601 state nodes and 400 control calls. H holds $(0,0,1)$ m; L spans 0.5 m, C has radius 0.25 m, and F8 semiaxes are 0.25 and 0.125 m. Maneuvers occupy 1--6 s with zero desired yaw. All methods share reference, initial state, plant, force, and limits within each protocol.

\begin{equation}\label{eq:metrics}\begin{aligned}e(t)&=\|\mathbf p(t)-\mathbf p_{\rm ref}(t)\|_2,\\J_{\rm IAE}&=\int_0^{8\,\mathrm s}e(t)\,dt,\\e_{\rm peak}&=\max_{0\le t\le8\,\mathrm s}e(t),\quad e_{\rm end}=e(8\,\mathrm s).\end{aligned}\end{equation}

IAE is trapezoidally integrated in m\,$\cdot$\,s. Current root-mean-square error (RMSE) integrates $e^2$ and divides by 8 s before taking a square root; the historical archive used a node average, but PID RMSE in Supplementary Table~\ref{tab:all_grid} is recomputed under the current integral definition. Terminal error is the 8 s value, not the final-second mean. These descriptive measures have different orderings and are not stability tests.

The policy benchmarks shown here use Seed 1 (6201), fixed before inspecting performance, and deterministic evaluation. Other trained seeds remain in the engineering evidence. Each uses 6400 interactions (16 complete eight-second episodes) and 6368 updates. The two-layer, 64-unit actor/critic, replay size 1600 and batch 32 are fixed finite-budget choices; the learning rate $3\times10^{-4}$, discount 0.99 and target update 0.005 follow the versioned SAC defaults \cite{r38}. Equal budgets permit a comparison of these deployed implementations, not a convergence claim. The network input and reward, including reference signals, are defined in Supplementary Section~\ref{sec:rl_details}.

For supervised fitting, 2400 training samples determine a 10--32--6 tanh model and input/output normalization. Still-air training separates intrinsic mismatch from the force applied during evaluation. Width 32 and the fitting budget limit computation and are project choices, rather than literature-optimal values. Residual augmentation and the model-capacity/runtime trade-off are motivated by learned quadrotor MPC \cite{r3,r7}. The independently tabulated 7200-sample prediction assessment uses a frozen no-margin criterion against zero residual; the criterion is reported even when it fails. Sampling, horizon and integrator settings are common controlled choices, with exact limits and rationale in Supplementary Table~\ref{tab:setting_basis}.

\noindent\begin{minipage}{\columnwidth}
\captionof{table}{Method names used in the paper. Suffixes describe deployment, not separately trained supervised models. Raw historical identifiers are mapped in the evidence package.}\label{tab:methods}
\centering\small\setlength{\tabcolsep}{3pt}
\begin{tabular}{lp{.67\columnwidth}}\toprule
ID & Method and defining change\\\midrule
E0 & PID; current-reference feedback\\
E1 & Nominal NMPC; no learned residual\\
E2 & \textbf{Supervised cyclewise / stagewise NMPC}; same frozen $g_\phi$\\
\bottomrule
\end{tabular}\end{minipage}

Table~\ref{tab:methods} defines the basic methods; Table~\ref{tab:rl_interfaces} separates the policy interfaces and their matched baselines. Historical PID trajectories are reused after the existing 220-case audit of plant/assets, references, force and initial state, and current performance/success rules. This supports matched trajectory comparisons; it does not identify the historical and current nominal implementations as identical. Historical percentages and PID timing remain separate (Supplementary Section~\ref{sec:pid_details}). The 220-case grid, the 32-case fixed-compensator confirmation, the random-force bank, and the CFD environment are analyzed separately. Their means describe finite prescribed conditions; cases, time nodes, paired records and training seeds are not interchangeable independent replications \cite{r37,r40}.
\subsection{Matched benchmarks and supervised residual performance}\label{sec:comparison}
\begin{figure*}[!t]
\centering
\begin{subfigure}[t]{.48\textwidth}
  \centering
  \includegraphics[width=\linewidth]{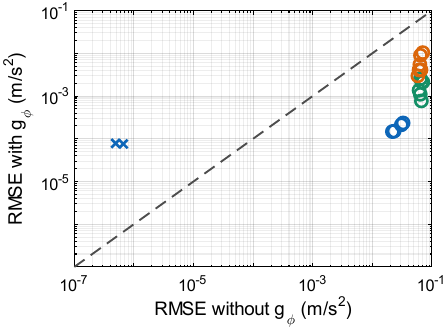}
  \caption{Body $x$ linear-acceleration residual}\label{fig:prediction_1}
\end{subfigure}\hfill\begin{subfigure}[t]{.48\textwidth}
  \centering
  \includegraphics[width=\linewidth]{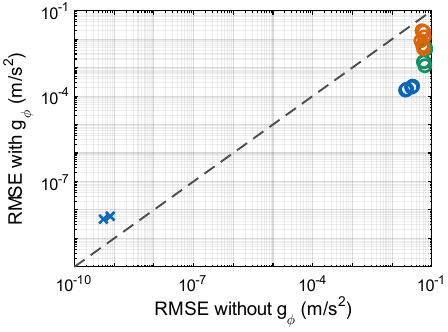}
  \caption{Body $y$ linear-acceleration residual}\label{fig:prediction_2}
\end{subfigure}

\predictionlegend
\caption{Prediction RMSE without and with frozen $g_\phi$ for body $x$ and $y$ linear-acceleration residuals (m/s$^2$). Each panel contains 18 groups, six per task. Colors identify L, C and F8; circles pass and crosses fail the prediction criterion. Below the diagonal learned prediction improves; above it prediction worsens. Most points improve, but two L points in each panel lie above the diagonal. Both axes are logarithmic with equal ranges within each panel. The six-component assessment and failed cases are retained in Supplementary Figure~\ref{fig:figS01_prediction} and Table~\ref{tab:prediction_fail}.}\label{fig:prediction_main}
\par\smallskip\noindent Figure~\ref{fig:prediction_main} shows broad prediction improvement with two L counterexamples per component. It therefore supports a useful but imperfect residual predictor; it does not establish a tracking benefit, which is tested separately in Figure~\ref{fig:fig02_absolute_grid}.
\end{figure*}

Figure~\ref{fig:prediction_main} assesses residual prediction independently of tracking. Fitting and normalization used only 2400 still-air training rows from the 4800-sample collection. A later independent assessment contains 18 groups and 7200 samples. All 18 pooled task--component comparisons beat or equal the zero-residual predictor, but only 100 of 108 finer group--component criteria pass. All eight failures occur in L. The original no-margin prediction gate remains failed (Supplementary Figure~\ref{fig:figS01_prediction}). Some failed errors are near numerical zero; others reach $7.56$--$7.78\times10^{-5}$ m/s$^2$ and $4.65$--$4.90\times10^{-4}$ rad/s$^2$. Small denominators can inflate ratios, but small absolute error does not retrospectively waive a frozen gate.

This comparison asks whether the same learned model improves executed tracking, and whether an additional policy supplies a further increment. Every curve reports its own error relative to the trajectory reference. H, L and F8 illustrate static and moving tasks; aggregate means retain all four tasks and directions. The nine displayed identities, prediction checks and remaining task histories are retained in the supplement, so the display selection does not determine the reported ranking.

\begin{figure*}[!t]
\centering
\begin{subfigure}[t]{.32\textwidth}
  \centering
  \includegraphics[width=\linewidth]{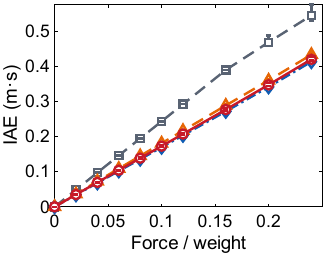}
  \caption{H: mean IAE}\label{fig:grid_H}
\end{subfigure}\hfill\begin{subfigure}[t]{.32\textwidth}
  \centering
  \includegraphics[width=\linewidth]{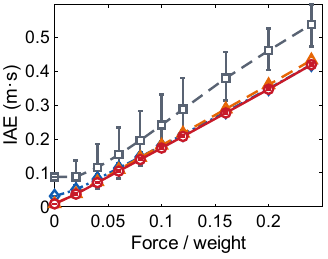}
  \caption{L: mean IAE}\label{fig:grid_L}
\end{subfigure}\hfill\begin{subfigure}[t]{.32\textwidth}
  \centering
  \includegraphics[width=\linewidth]{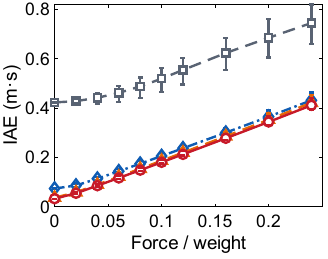}
  \caption{F8: mean IAE}\label{fig:grid_F8}
\end{subfigure}
\par\smallskip
\begin{subfigure}[t]{.32\textwidth}
  \centering
  \includegraphics[width=\linewidth]{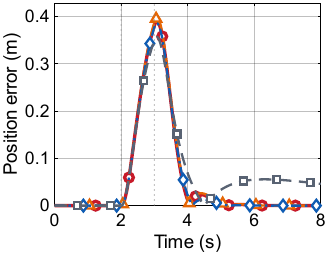}
  \caption{H: reference error}\label{fig:history_H}
\end{subfigure}\hfill\begin{subfigure}[t]{.32\textwidth}
  \centering
  \includegraphics[width=\linewidth]{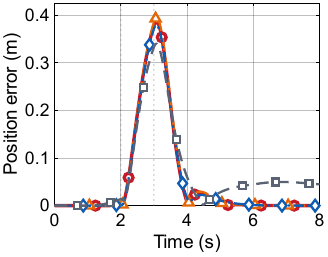}
  \caption{L: reference error}\label{fig:history_L}
\end{subfigure}\hfill\begin{subfigure}[t]{.32\textwidth}
  \centering
  \includegraphics[width=\linewidth]{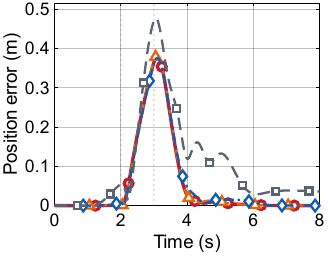}
  \caption{F8: reference error}\label{fig:history_F8}
\end{subfigure}
\basiclegend
\caption{Absolute tracking under the pulse protocol. Upper panels average six directions at each positive force/weight ratio; bars span direction minima--maxima, and zero has one case. Lower panels retain every state node; sparse, staggered markers distinguish overlapping curves. They use the fixed $\lambda=0.24$, horizontal $+45^\circ$ cases; vertical dotted lines mark force onset and removal. This figure shows a moving-task recovery benefit from supervision, while H remains a counterexample. Every method is compared with its own trajectory reference; PID uses the documented implementation bridge. All four tasks enter the aggregate means.}\label{fig:fig02_absolute_grid}
\end{figure*}

Figure~\ref{fig:fig02_absolute_grid} tests the two deployment rules under identical pulse loads. The upper panels average directions, whereas panels~\subref{fig:history_H}--\subref{fig:history_F8} locate the error in time for one fixed direction. PID has larger IAE in these examples, yet its H transient peak need not be larger. NMPC combines preview, a dynamics model and optimization; this comparison does not isolate any one of those mechanisms. Historical PID--historical nominal percentages are not inferred from the current nominal curves.

\begin{figure*}[!t]\centering
\begin{subfigure}[t]{.32\textwidth}\centering
\includegraphics[width=\linewidth]{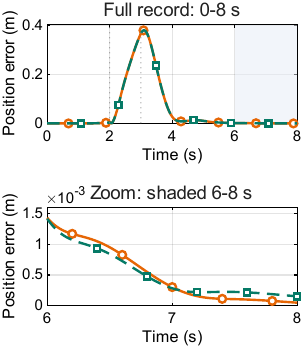}
\caption{External RL}\label{fig:rl_1_F8}
\end{subfigure}\hfill\begin{subfigure}[t]{.32\textwidth}\centering
\includegraphics[width=\linewidth]{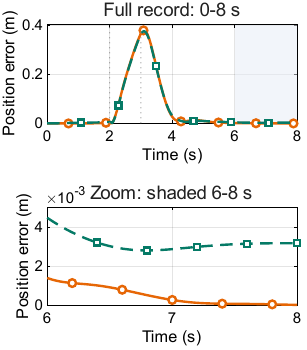}
\caption{Internal cyclewise RL}\label{fig:rl_2_F8}
\end{subfigure}\hfill\begin{subfigure}[t]{.32\textwidth}\centering
\includegraphics[width=\linewidth]{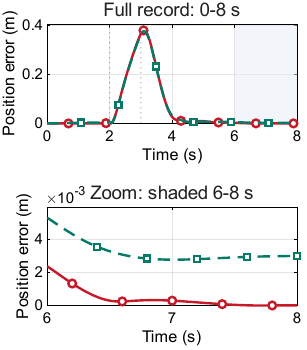}
\caption{Internal stagewise RL}\label{fig:rl_3_F8}
\end{subfigure}
\rllegend
\caption{F8 policy comparisons at $\lambda=0.24$, horizontal $+45^\circ$, using Seed 1 (6201). Upper plots show 0--8 s; lower plots enlarge the shaded 6--8 s interval of the same error, also in m. Warm curves are matched supervised baselines. External RL remains close to its baseline; internal RL leaves larger tail errors. H/L histories are in Figure~\ref{fig:rl_other_histories}; Table~\ref{tab:rl_main} reports the full grid.}\label{fig:fig04_rl_absolute}
\par\smallskip\noindent\begin{minipage}{\textwidth}
In this F8 case, external RL remains close to its supervised baseline, whereas both internal policies leave larger errors over 6--8 s and impair late recovery. The nearly overlapping full-record peaks therefore conceal a late-tracking difference; the finite tail window does not establish asymptotic steady-state error.
\end{minipage}
\end{figure*}

The supervised differences depend on task and phase. H remains worse than nominal, including all 55 conditions. On moving tasks, close forcing peaks can coexist with better recovery: a smaller area after removal lowers full-window IAE without a smaller maximum deviation. Across all 220 cases, cyclewise and stagewise supervision reduce nominal mean IAE by 4.38\% and 7.77\%, and stagewise beats cyclewise in 217 cases (Supplementary Table~\ref{tab:all_grid}). Cyclewise holds a current-state correction; stagewise samples predicted node states before solving. Their different treatment of predicted motion is consistent with the moving-task gain, but these tests do not isolate it as the unique cause. The supported choice is stagewise deployment for the measured mean tracking objective, with hover and forcing-peak limitations retained. Figure~\ref{fig:fig04_rl_absolute} next tests whether a policy improves its corresponding supervised baseline.

\begin{table}[!t]\caption{Policy benchmarks test an increment over supervision; they are not required by the proposed method.}\label{tab:rl_interfaces}\centering\small
\begin{tabular}{lp{.52\columnwidth}}\toprule
ID & Method and defining change\\\midrule
E3: external RL & Four motor actions after the solve; cyclewise supervision\\
E4: internal cyclewise & Six acceleration actions before the solve; cyclewise supervision\\
E4: internal stagewise & Six cycle-level acceleration actions; stagewise supervision\\\bottomrule
\end{tabular}\end{table}

\begin{table}[!t]\caption{Complete 220-case means for the main Seed 1 policy comparisons. The appropriate supervised mean is lower for each interface; all policy rows use Seed 1 (6201).}\label{tab:rl_main}\centering\small\begin{tabular}{lr}\toprule Method & IAE (m\,$\cdot$\,s)\\\midrule
Supervised cyclewise & \textbf{0.207016}\\
External RL & 0.207266\\
Internal cyclewise RL & 0.220089\\\midrule
Supervised stagewise & \textbf{0.199679}\\
Internal stagewise RL & 0.212478\\
\bottomrule\end{tabular}\end{table}
Table~\ref{tab:rl_main} separates the cyclewise and stagewise families. External RL increases cyclewise mean IAE from 0.207016 to 0.207266 m\,$\cdot$\,s; only 3 of 220 cases improve. Its terminal mean rises from 0.038 to 0.130 mm. A small force-window gain is outweighed by pre- and post-force losses. Post-solve injection changes the executed input without changing the solved plan. This verified interface mismatch is consistent with a loss, but is not isolated as its sole cause.

Internal cyclewise and stagewise means are 0.220089 and 0.212478 m\,$\cdot$\,s, above their matched supervised means. Some peak and RMSE measures improve while terminal errors reach 3.178 and 3.017 mm, respectively. Both internal deployments worsen all four no-force controls, so large applied forces alone cannot explain the loss. Holding the actor fixed and switching to stagewise evaluation improves all 220 cases, but the no-RL model benefits from the same switch. This improvement belongs to the deployment rule, not automatically to RL.

Figure~\ref{fig:fig04_rl_absolute} shows that the three policies can have nearly identical forcing peaks yet different late tracking. In panel~\subref{fig:rl_1_F8}, external RL stays close to its cyclewise baseline: their 6--8 s mean errors are $5.079\times10^{-4}$ and $5.042\times10^{-4}$ m. Panels~\subref{fig:rl_2_F8},\subref{fig:rl_3_F8} show larger internal-policy offsets, with means $3.184\times10^{-3}$ and $3.245\times10^{-3}$ m, versus $5.042\times10^{-4}$ and $4.350\times10^{-4}$ m for the matched baselines. Thus these internal corrections impair late recovery in this F8 case; close full-scale curves do not imply equal performance. This finite window does not establish an asymptotic steady-state error. Both rows show the same $e(t)=\|\mathbf p-\mathbf p_{\rm ref}\|_2$ in m, without rescaling the data, smoothing, or a new run. Table~\ref{tab:rl_main} determines the complete-grid conclusion. Supplementary Figure~\ref{fig:rl_other_histories} retains H/L; Table~\ref{tab:absolute_windows} and Figure~\ref{fig:rl_intensity_full} report the window summaries and intensity dependence.

Training return rises between the first and final four-episode cycles, predominantly through reduced action-magnitude and action-change costs in the external branch. Improved return does not establish improved tracking. The discounted squared-error/action objective differs from the undiscounted linear-error IAE; $\gamma=0.99$ and $\Delta=0.02$ s give an e-fold time of about 1.99 s. SAC optimizes return and entropy \cite{r28,r35}; it does not guarantee an increment over this supervised baseline. The external package has lower IAE than internal cyclewise in all 220 cases, but differs in action authority, dimension and observation. These are package-specific comparisons under a fixed budget, not a general ranking of RL interfaces.

The supervised comparison is therefore the central method result, with RL supplying an incremental benchmark rather than an assumed performance ladder.

Mirror equivariance does not enforce every missing zero-output condition: only components reversing sign are zero at mirror-fixed inputs. Forward acceleration and pitch acceleration do not reverse under this mirror. This algebra explains why symmetry compliance and componentwise qualification can disagree; it does not identify training coverage or approximation as the unique source of the remaining error.

On the matched 220 cases, nominal, cyclewise, and stagewise mean IAE are 0.216494, 0.207016, and 0.199679 m\,$\cdot$\,s. Cyclewise and stagewise corrections improve 135 and 155 cases relative to nominal; stagewise beats cyclewise in 217 of 220. Both worsen all 55 hover cases, while moving-task and no-force controls support compensation of intrinsic motion-dependent mismatch. The cyclewise RMSE and peak rise to 0.056876 and 0.183818 m from nominal 0.056253 and 0.178509 m. Stagewise RMSE falls to 0.055245 m, but peak remains slightly above nominal at 0.178996 m.

Temporal decomposition makes the gain interpretable. Relative to nominal, the mean force-window IAE rises by 0.00189787 and 0.00025443 m\,$\cdot$\,s for cyclewise and stagewise rules, while the 3--8 s area falls by 0.01097907 and 0.01637237 m\,$\cdot$\,s. The net gain is principally post-force, not a uniformly smaller force transient. Cyclewise repeats one current-state residual; stagewise samples shifted prediction nodes before solving and fixes that sequence during optimization. Improved predictive fidelity is a plausible mediator, not an isolated cause. Existing own-command replay grows from about 0.218 mm one-step position discrepancy to 82.90--82.94 mm at 20 steps from force onset. It is not a common-input causal comparison. Prediction-fit and feedback objectives differ \cite{r39}; their coexistence here is established by separate tests.

Taken together, these results support stagewise supervision as the main tested deployment of this frozen residual: it has lower mean IAE than cyclewise supervision, improves most matched cases, and requires no added policy training. They also define its limits. Neither the detailed prediction gate nor all-task improvement is achieved, and the aggregate gain is mostly recovery rather than uniformly smaller forcing peaks. This is a bounded method choice for the measured tracking objective, not a claim of superiority over every compensated controller.
\subsection{Additional fixed compensation: tracking gain and angular cost}\label{sec:fixed_main}

This comparison asks whether online information from the preceding transition adds value beyond frozen supervision, and what other motion it costs. Each fixed-gain variant retains the stagewise $g_\phi$ model and adds causal innovation compensation to that baseline. Cyclewise supervision is a separate same-model deployment reference in the four-method comparison. The 32-case confirmation uses newly prescribed scenes after development selection; it is independent of the 220-case summary and does not confirm any RL recipe.

Fixed compensation uses $\beta=0.2$ and selects $\kappa=1.5$ on the 24-case development set (D24) after limited gain comparisons. T32 then freezes four methods on 32 separately generated cases, with actual force onsets/durations different from the 220 grid. Mean IAE is 0.172596, 0.166359, 0.123642 and 0.120661 m\,$\cdot$\,s for supervised cyclewise, stagewise and gains 1, 1.5. Gain 1.5 beats stagewise IAE in all 32 cases, reducing its mean by 27.47\%. The mean paired gain-1.5-minus-stagewise difference is $-0.045698$ m\,$\cdot$\,s; its 95\% stratified descriptive resampling interval is $[-0.049957,-0.041606]$ m\,$\cdot$\,s (Supplementary Table~\ref{tab:t32_intervals}). This confirms the selected fixed compensator on this bank, not a trained RL policy or a population-wide effect.

The stronger comparator also costs angular motion: its mean angular-rate RMSE is 0.118341 rad/s versus stagewise 0.107402 rad/s, and mean attitude RMSE is slightly higher. Direct innovation compensation is consistent with correcting observed short-term mismatch \cite{r41}; its acceleration estimate includes model and discretization error. Event-window secondary analysis uses actual onset/removal times after correcting an inherited fixed-window script. Timing boundaries differ across T32 implementations, so they do not support an overhead ranking. Full identity-enumeration and receipt corrections are disclosed in the supplement.

\noindent\begin{minipage}{\columnwidth}
\captionof{table}{Separate 32-case confirmation: gain 1.5 lowers position IAE but raises angular-rate RMSE relative to stagewise supervision. Absolute case means cover four frozen methods, all completing eight seconds with zero fallback. Bold identifies the lowest value in each matched column; units and protocols are not pooled with the 220-case grid.}\label{tab:t32_main}
\centering\small\setlength{\tabcolsep}{3pt}
\begin{tabular}{lrrr}\toprule
Method & IAE (m\,$\cdot$\,s) & Att. (rad) & Rate (rad/s)\\\midrule
Supervised cycle & 0.172596 & 0.041300 & 0.107403\\
Supervised stage & 0.166359 & 0.041076 & \textbf{0.107402}\\
Fixed gain 1 & 0.123642 & \textbf{0.041030} & 0.116129\\
Fixed gain 1.5 & \textbf{0.120661} & 0.041335 & 0.118341\\\bottomrule
\end{tabular}\end{minipage}

Table~\ref{tab:t32_main} makes the trade-off explicit: gain 1.5 lowers mean position IAE by 27.47\%, but increases mean angular-rate RMSE by 10.19\% and attitude RMSE by about 0.63\% relative to stagewise supervision. Gain 1 slightly improves attitude RMSE instead. The rate metric is an error relative to the body-coordinate reference, not a frequency-domain measure of oscillation or a stability certificate. Supplementary Figure~\ref{fig:activity} locates the angular-rate and command-increment excursions in time; the all-case summaries below quantify their size. These checks test input activity, rather than assume that larger RMSE means high-frequency vibration. Across the same 32 scenes, the mean case RMS of consecutive four-motor command increments increases from 6.726 to 10.749 rad/s (59.80\%) for gain 1.5, consistent with greater input activity. These increments include onset and removal transients; they do not alone establish high-frequency jitter or instability.

The causal innovation updates a mismatch estimate from the last executed transition, whereas the frozen supervised model learns still-air acceleration differences. Their information sources are complementary. Disturbance-observer theory motivates feeding back this model discrepancy \cite{r41}; the present matched test demonstrates the position benefit of the selected implementation and its secondary cost, not identification of the true force. More active tilt/rate correction is consistent with faster translational recovery, but its unique physical cause has not been isolated. For a position-priority task this compensator is the stronger tested comparator; for reduced angular-rate error, supervision is preferable on this bank. Without a prespecified acceptable angular-motion limit, neither observation supports a universal combined ranking.
\subsection{Validation under prescribed random forces}

\begin{figure*}[!t]
\centering
\begin{subfigure}[t]{.32\textwidth}
  \centering
  \includegraphics[width=\linewidth]{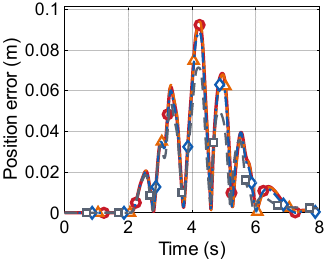}
  \caption{H: fixed random-force case}\label{fig:wn_H}
\end{subfigure}\hfill\begin{subfigure}[t]{.32\textwidth}
  \centering
  \includegraphics[width=\linewidth]{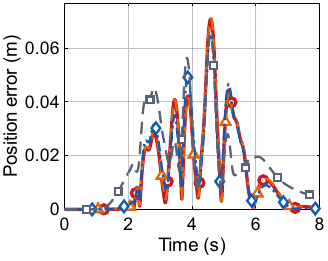}
  \caption{L: fixed random-force case}\label{fig:wn_L}
\end{subfigure}\hfill\begin{subfigure}[t]{.32\textwidth}
  \centering
  \includegraphics[width=\linewidth]{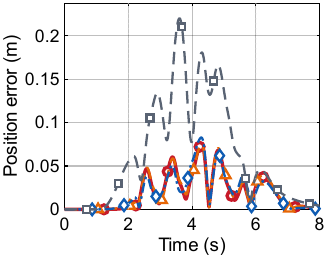}
  \caption{F8: fixed random-force case}\label{fig:wn_F8}
\end{subfigure}
\basiclegend
\caption{Absolute reference-error histories for the frozen band-limited random-force protocol. Selection is fixed at RMS force/weight 0.10, horizontal $+45^\circ$, realization 00; the same case rule is used for all three tasks. Dotted lines bound the 2--6 s activation window. This figure illustrates task-dependent tracking and late error; Table~\ref{tab:wn_main} and Supplementary Table~\ref{tab:wn_all} report all 96 blocks, so no ranking is inferred solely from these three histories. WN is a project identifier, not white noise or atmospheric validation.}\label{fig:fig06_wn_absolute}
\end{figure*}
Figure~\ref{fig:fig06_wn_absolute} asks whether the pulse findings depend on a single rectangular forcing episode. Each panel plots position error against its own reference, showing several forced excursions and a later return. In H, both supervised rules raise IAE and peak error relative to nominal. In L, stagewise slightly lowers IAE despite a larger peak. In F8, stagewise lowers both measures, whereas cyclewise lowers the peak but raises IAE. Thus even these fixed examples show why a smaller peak cannot substitute for the integrated tracking result. WN is the project identifier for this frozen band-limited random-force protocol; it does not denote white noise. It tests transfer from a rectangular pulse to a time-varying load without changing actors. The frozen scalar force sums harmonics from 0.5 to 5 Hz with deterministic hashed phases, tapered activation over 2--6 s, and active-window RMS normalization. World directions X/Y/$+45^\circ$, force-RMS/weight ratios 0.05 and 0.10, four realizations, and four tasks produce 96 paired environment blocks. Additional structural identities use a prescribed 16-block subset, reported separately in the supplement. The frozen protocol has terminal verification; all admitted evaluations complete eight seconds with zero fallback. A plant audit record is a check of a run, not an additional experiment.

\begin{table}[!t]\caption{All 96 random-force blocks, with Seed 1 shown for the policy comparisons. Nominal and each matched supervised baseline remain distinct; all policy rows use Seed 1 (6201).}\label{tab:wn_main}\centering\small\begin{tabular}{lr}\toprule Method & IAE (m\,$\cdot$\,s)\\\midrule
Nominal NMPC & 0.107246\\
Supervised cyclewise & \textbf{0.097677}\\
External RL & 0.097897\\
Internal cyclewise RL & 0.106968\\\midrule
Supervised stagewise & \textbf{0.094390}\\
Internal stagewise RL & 0.103561\\
\bottomrule\end{tabular}\end{table}
Nominal mean IAE is 0.107246 m\,$\cdot$\,s. Cyclewise and stagewise supervised means are 0.097677 and 0.094390 m\,$\cdot$\,s, with 61 and 66 improvements against nominal. Stagewise improves 94 of 96 blocks against cyclewise. This reproduces the direction of the supervised aggregate benefit under another prescribed load process, while retaining nonuniform case behavior. It does not show that the residual estimates WN, because it was trained without that force and does not receive its realization.

The external RL mean is 0.097897 m\,$\cdot$\,s, above cyclewise supervision, with one improvement among 96 blocks. Internal cyclewise and stagewise means are 0.106968 and 0.103561 m\,$\cdot$\,s, also worse than their matched supervised means. V2 improves its original internal-cyclewise policy on all 16 structural blocks for this seed. This subset is reported separately (Supplementary Table~\ref{tab:wn_all}); comparisons to supervision must use those same blocks, not the 96-block mean.

The complete-block comparison, illustrated in Figure~\ref{fig:fig06_wn_absolute}\,\subref{fig:wn_H}--\subref{fig:wn_F8}, weakens an explanation based solely on the original pulse waveform: the negative mean RL increment persists after changing the load process. It does not identify a universal RL limitation or the dominant learned failure mechanism. The fixed force bank, paired blocks, and unchanged weights support this transfer statement; ideal sensing, limited strengths and a synthetic process bound it. Four realizations per prescribed condition support within-protocol variability reporting, not unrestricted atmospheric inference. Controller seeds and time samples are not independent environment replications \cite{r37}.

\subsection{Supplementary validation in a synthetic CFD environment}\label{sec:cfd_results}
This comparison changes the rotor-inflow environment while freezing controllers. All five methods share the field, initial state, reference, plant, motor bounds and eight-second duration; the NMPC methods also share solver settings. The zero-residual control (N-off) retains the stagewise prediction graph with $g_\phi=0$; it is not identified with historical nominal NMPC. Cyclewise and stagewise supervision use the frozen network; the internal stagewise policy uses the retained Seed 1 actor, without a horizon-decay intervention. Neither network receives CFD features. PID supplies a feedback benchmark. Twenty completed runs comprise five methods on four references. The 40 derived paired rows and independent plant audits verify or compare those same runs, rather than add performance samples.

OpenFOAM v2512 evolves a synthetic decaying field in a periodic 8 m cube on $64^3$ cells with wall-adapting local eddy-viscosity (WALE) large-eddy simulation (LES) \cite{r46,r47}. The mean is $(3,0,0)$ m/s; the initial vector fluctuation RMS of 0.75 m/s decays to about 0.310 m/s without sustained forcing. Velocity is interpolated at each rotor and mapped through the existing aerodynamic plant. This is neither aircraft surface-pressure CFD nor a resolved rotor-wake calculation. Supplementary Tables~\ref{tab:cfd_settings}, \ref{tab:cfd_position} and \ref{tab:cfd_secondary} give the exact settings, position outcomes and secondary costs.
\begin{figure*}[!t]
\centering
\begin{subfigure}[t]{.48\textwidth}
\centering\includegraphics[width=\linewidth]{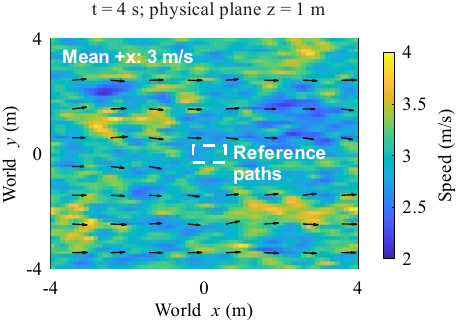}
\caption{Velocity magnitude, physical plane $z=1$ m}\label{fig:cfd_plane}
\end{subfigure}\hfill
\begin{subfigure}[t]{.48\textwidth}
\centering\includegraphics[width=\linewidth]{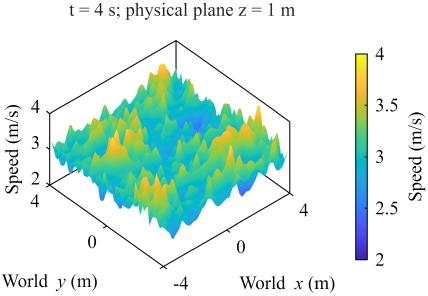}
\caption{Speed surface, not physical height}\label{fig:cfd_surface}
\end{subfigure}\par\smallskip
\begin{subfigure}[t]{.48\textwidth}
\centering\includegraphics[width=\linewidth]{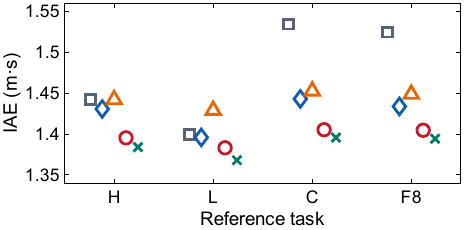}
\caption{Four tasks, five methods}\label{fig:cfd_iae}
\end{subfigure}\hfill
\begin{subfigure}[t]{.48\textwidth}
\centering\includegraphics[width=\linewidth]{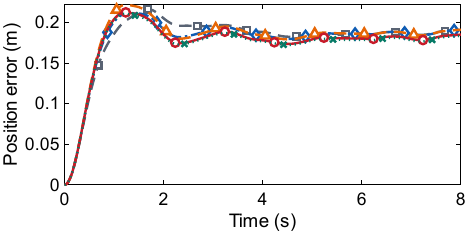}
\caption{H: continuous CFD inflow}\label{fig:cfd_H}
\end{subfigure}\par\smallskip
\begin{subfigure}[t]{.48\textwidth}
\centering\includegraphics[width=\linewidth]{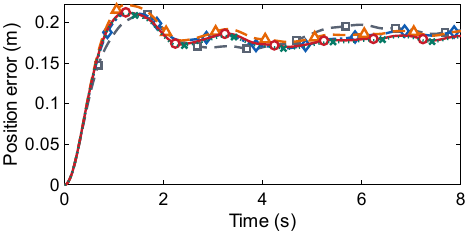}
\caption{L: continuous CFD inflow}\label{fig:cfd_L}
\end{subfigure}\hfill
\begin{subfigure}[t]{.48\textwidth}
\centering\includegraphics[width=\linewidth]{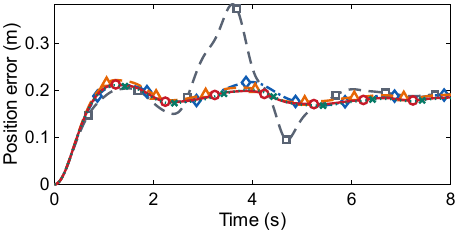}
\caption{F8: continuous CFD inflow}\label{fig:cfd_F8}
\end{subfigure}\par\smallskip
\cfdlegend
\caption{One prescribed CFD field. (a,b) The same $t=4$ s, $z=1$ m slice is shown as a plane and speed surface; surface height is speed, not elevation. Black arrows show local horizontal velocity; the mean flow is $+x$. The dashed box marks the reference-path region. Both views use the same 2--4 m/s color scale. (c) Absolute IAE for all 20 runs; markers are offset by method. (d)--(f) Full unsmoothed histories. Stagewise supervision improves the zero-residual control; the retained RL policy adds a small position benefit with higher angular-rate RMSE.}\label{fig:cfd}
\end{figure*}
Four-task mean IAE is 1.475109 m\,$\cdot$\,s for PID, 1.425794 for the zero-residual control, 1.443159 for cyclewise supervision, 1.397113 for stagewise supervision and 1.385507 for internal stagewise RL. Stagewise supervision reduces the zero-residual mean by 2.01\% and improves each task; cyclewise supervision does not improve that mean. The stagewise benefit coexists with slightly higher H/L peaks, velocity and attitude errors. All 20 runs complete with zero fallback. These observations support stagewise deployment in this prescribed field, rather than uniform superiority of either supervised rule.

The retained RL actor further reduces stagewise position IAE by 0.83\% and improves all four cases, but angular-rate RMSE increases in every task (four-task means 0.095271 versus 0.094892 rad/s). Thus the CFD result is a small positive position increment with an angular-motion cost. It does not contradict the negative mean increments on the pulse and random-force banks: different load processes and a different matched computational control define different tests. No second CFD seed was evaluated, and the actor was not selected by this field result.

Figure~\ref{fig:cfd}\,\subref{fig:cfd_plane},\subref{fig:cfd_surface} shows spatial speed variation around the imposed mean flow, rather than a uniform inflow: all 4096 field samples are used, and local arrows retain transverse velocity rather than being forced parallel to the mean. Panel~\subref{fig:cfd_iae} compares the 20 complete-run IAEs; panels~\subref{fig:cfd_H}--\subref{fig:cfd_F8} locate differences and persistent offsets in time. The boxed reference region is a path bound, not an aircraft footprint. The H/L histories show closely spaced controller responses under sustained loading; in F8 the larger PID transient is visible. The stagewise and retained RL gains reduce integrated error without eliminating the persistent offset. These views separate the applied environment from controller performance. Fixed 0--2, 2--4 and 4--8 s windows locate the stagewise benefit without importing pulse recovery. Stagewise IAE is lower than the zero-residual control in both later windows for all four tasks; persistent loading leaves a late position offset. The learned correction acts on prediction without access to the realized flow. This is evidence of limited closed-loop transfer, not turbulent-force identification. The changed hover ordering cannot be attributed to turbulence alone because the load map and matched nominal graph also change. One field and four reference trajectories do not support atmospheric generalization or a population confidence interval.

\subsection{Reliability and measured computation}

\begin{figure}[!t]
\centering\includegraphics[width=\columnwidth]{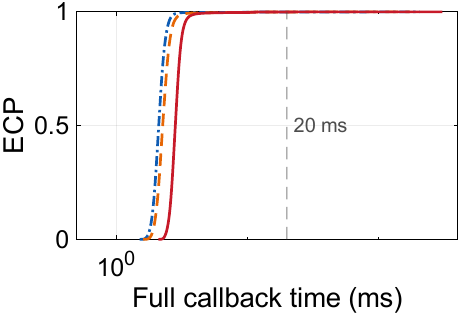}
\par\smallskip{\small\keyline{nomcolor}{Nominal}\keyline{cyclecolor}{Cyclewise}\keyline{stagecolor}{Stagewise}}
\caption{Measured full-callback empirical cumulative probabilities (ECP) for nominal and the two supervised rules: 88,000 calls per identity. P95 is 2.505, 2.733 and 3.338 ms, respectively. Only the 20 ms control period is marked; the separate 50 ms screening rule remains in the supplement. The figure shows a modest increase in typical callback time from nominal to stagewise deployment, with rare exceedances. These separately acquired distributions do not establish a hard-real-time guarantee.}\label{fig:fig08_callback_cdf}
\end{figure}

Figure~\ref{fig:fig08_callback_cdf} tests computation rather than tracking: at any horizontal time threshold its ordinate is the fraction of measured full callbacks at or below that threshold. A rightward curve requires more time at the same probability. The supervised rules increase typical callback time relative to nominal NMPC, and stagewise is slower than cyclewise; all three P95 values remain below 20 ms, but the tail prevents a hard-real-time claim. All original current-grid identities complete their prescribed trajectories, but measured deadline exceedances remain a separate result. Full callback includes preparation, network inference when inside the recorded boundary, parameter assembly, solve, retrieval, and applied-command update. Supplementary Table~\ref{tab:timing} gives pooled quantiles and exceedance counts, not an average of case P95 values. Stagewise sampling and policy inference add implemented operations; separate timing batches do not isolate their causal cost from system load. The simulation is synchronous: elapsed wall time is not injected as plant delay, so missed physical deadlines cannot explain its tracking loss.

The stagewise callback evaluates 20 stage corrections, versus one cyclewise evaluation; each sequence is fixed during the one SQP-RTI update. The observed P95 increase from 2.733 to 3.338 ms is consistent with this added preparation, but separate acquisitions and host load prevent isolating inference as its sole cause. These quantiles lie below the 20 ms period, while the stagewise maximum reaches 305.087 ms. Thus the supervised method is computationally feasible in the measured synchronous simulation, with rare deadline exceedances that require separate treatment in physical deployment. Original-policy timing summaries are tabulated separately, without pooling their samples or substituting a mean case P95 for the pooled quantile.

Supplementary Figure~\ref{fig:other_tasks}\,\subref{fig:other_C_pulse},\subref{fig:cfd_other} completes the C histories under the pulse and CFD protocols. The fixed-compensation motion and input checks are in Figure~\ref{fig:activity}. Supplementary Table~\ref{tab:seed_timing} reports Seed 1 per-case timing summaries separately from pooled callback quantiles.

\section{Discussion}\label{discussion}

The deployment results distinguish integrated tracking benefit from disturbance attenuation. On the 220-case pulse bank, stagewise supervision lowers mean position IAE relative to nominal NMPC, but all 55 hover cases worsen and the reduction is dominated by the post-force period. The selected causal innovation compensator produces a larger position gain than stagewise supervision on a separate 32-case confirmation set, while increasing angular-rate RMSE. Accordingly, no controller is uniformly preferable without specifying task and secondary-motion constraints.

The supervised rules do not uniformly lower the force-window error or peak; their net improvement is mainly accumulated after removal. All 55 hover cases worsen, including the no-force control. Still-air training and the absence of force input are consistent with correction of intrinsic motion-dependent mismatch, rather than direct estimation of the imposed load. Small nonzero residuals near hover could matter when nominal errors are already small, but their causal role has not been isolated by a zeroing or re-fitting intervention. The failed prediction components likewise do not uniquely explain the hover behavior. Control-oriented model learning distinguishes fit from feedback performance \cite{r39}; here separate prediction and closed-loop assessments make that distinction concrete without asserting a causal relation from aggregate scores alone.

The policy benchmarks test an additional contribution on top of supervision. The retained external and internal policies have higher mean IAE than their matched supervised controls, both for pulses and random forces. External losses are small and coexist with occasional transient gains. The original reward decomposition explains why return can improve without reducing tracking IAE: most external return growth comes from action costs, and a discounted weighted squared objective differs from the reported undiscounted integral. It does not prove that reward weights alone caused the loss. The short finite budget, training-bank coverage and single displayed seed remain limitations of this comparison.

Attenuating the retained internal action along the horizon improves its parent policy; suppressing angular output alone changes little (Supplementary Table~\ref{tab:structural}). This intervention tests how a fixed actor correction is used, rather than a new training method. The result supports reduced persistence for this actor, while leaving its mean IAE above cyclewise supervision. It does not establish a unique learned failure mechanism.

On the separate confirmation bank, causal innovation feedback lowers position IAE but increases angular-rate error. Stagewise supervision is preferable for the latter metric; gain 1.5 is preferable for position IAE. Neither establishes a universal combined ranking without a predefined joint objective or angular limit. This non-RL comparator therefore remains part of the main-text comparison.

The five-method CFD comparison adds a narrow environment-transfer result: stagewise supervision lowers IAE on all four prescribed references without wind information or retraining, while H/L peaks and two other error measures slightly worsen. The retained internal stagewise actor additionally lowers position IAE by 0.83\%, with higher angular-rate RMSE in all four tasks. This protocol-specific positive increment prevents extending the original negative RL averages to every environment. Persistent offsets show that the learned still-air residual does not eliminate continuous wind loading. WALE supplies a modeled subgrid stress, not a validation of this field as atmospheric turbulence \cite{r46,r47}. There is no grid-convergence study or physical wind calibration, and the different hover ordering across protocols does not isolate a turbulence mechanism.

The evidence remains simulation with ideal state feedback, one numerical vehicle, finite force banks and one synthetic CFD field. Incomplete direction coverage, one displayed policy seed and disclosed development exposure limit generalization. Reused baselines and time samples do not increase replication. Callback measurements establish measured simulation cost, without validating physical stability or hard-real-time deployment.

\section{Conclusion}\label{conclusion}

A frozen supervised acceleration residual provides a useful, task-dependent augmentation to nonlinear predictive control of the evaluated force-disturbed quadrotor. Cyclewise and stagewise deployment use the same model but answer different prediction-horizon choices. Over the complete 220-condition grid, they lower mean position IAE by 4.38\% and 7.77\%, respectively; stagewise improves 217 of 220 cases against cyclewise. Most net benefit occurs after force removal. Both worsen all 55 pulse-grid hover conditions, and eight of 108 detailed prediction criteria remain failed. The supported conclusion is improved average executed tracking with a clear hover and transient limitation, rather than uniform disturbance rejection or fully qualified prediction.

The original external and internal RL benchmarks supply no additional mean IAE benefit over their respective supervised controls for the retained policies on the pulse grid. The observation persists under the separate prescribed random-force protocol, where cyclewise and stagewise supervision reduce the nominal mean by 8.92\% and 11.99\%. Thus the tested supervised benefit transfers across these two force processes, while a useful residual model does not automatically create a useful added policy. Structural interventions in the supplement test how the fixed internal correction is used; they do not establish a unique cause or exclude successful RL under other objectives, budgets and interfaces.

Additional fixed causal compensation answers a separate controller-selection question. On the 32-condition confirmation bank, gain 1.5 reduces mean position IAE by 27.47\% relative to stagewise supervision and improves every case, while increasing angular-rate RMSE by 10.19\% and slightly increasing attitude RMSE. It is consequently a stronger position-tracking comparator with a measured secondary-motion cost, not a low-performance method. Stagewise supervision is the main learned-model deployment supported by this study; a preference for it over fixed compensation must specify angular-motion priorities rather than disregard the lower position error.

A plant-only synthetic CFD field gives further, limited support across five methods: the frozen stagewise model lowers IAE in four matched trajectories, with a 2.01\% reduction of their mean. This comparison retains a same-graph zero-residual control and exposes secondary losses and sustained position offsets. Cyclewise supervision does not improve the zero-residual mean in this field. The retained internal stagewise policy further reduces stagewise position IAE by 0.83\%, but increases angular-rate RMSE in all four tasks. This bounded positive increment complements, rather than replaces, the original negative means. The comparison supports transfer without retraining to one prescribed field, rather than general turbulent-wind rejection.

Finally, the measured stagewise callback P95 of 3.338 ms is below the 20 ms control period, but rare exceedances preclude a hard-real-time claim. Within the tested simulation, the practical choice is therefore a supervised model whose horizon evaluation is checked against execution, with explicit attention to hover, recovery, angular motion and runtime. More elaborate corrections must demonstrate an increment over their matched reference.

\clearpage
\bibliographystyle{unsrtnat}
\begingroup\small\bibliography{references}\endgroup

@article{r1,
  title = {Differential Flatness of Quadrotor Dynamics Subject to Rotor Drag for Accurate Tracking of High-Speed Trajectories},
  author = {Faessler, Matthias and Franchi, Antonio and Scaramuzza, Davide},
  year = {2018},
  doi = {10.1109/LRA.2017.2776353},
  journal = {IEEE Robotics and Automation Letters},
  volume = {3},
  number = {2},
  pages = {620--626}
}

@article{r2,
  title = {On Trajectory Tracking Model Predictive Control of an Unmanned Quadrotor Helicopter Subject to Aerodynamic Disturbances},
  author = {Alexis, K. and Nikolakopoulos, G. and Tzes, A.},
  year = {2014},
  doi = {10.1002/asjc.587},
  journal = {Asian Journal of Control},
  volume = {16},
  number = {1},
  pages = {209--224}
}

@article{r3,
  title = {Data-Driven {MPC} for Quadrotors},
  author = {Torrente, Guillem and Kaufmann, Elia and Föhn, Philipp and Scaramuzza, Davide},
  year = {2021},
  doi = {10.1109/LRA.2021.3061307},
  journal = {IEEE Robotics and Automation Letters},
  volume = {6},
  number = {2},
  pages = {3769--3776}
}

@book{r4,
  title = {Nonlinear Model Predictive Control},
  author = {Grüne, Lars and Pannek, Jürgen},
  year = {2017},
  doi = {10.1007/978-3-319-46024-6},
  booktitle = {Communications and Control Engineering},
  publisher = {Springer International Publishing}
}

@inproceedings{r5,
  title = {A model predictive controller for quadrocopter state interception},
  author = {Mueller, Mark W. and D'Andrea, Raffaello},
  year = {2013},
  doi = {10.23919/ECC.2013.6669415},
  booktitle = {2013 European Control Conference (ECC)},
  pages = {1383--1389}
}

@article{r6,
  title = {Performance, Precision, and Payloads: Adaptive Nonlinear {MPC} for Quadrotors},
  author = {Hanover, Drew and Föhn, Philipp and Sun, Sihao and Kaufmann, Elia and Scaramuzza, Davide},
  year = {2022},
  doi = {10.1109/LRA.2021.3131690},
  journal = {IEEE Robotics and Automation Letters},
  volume = {7},
  number = {2},
  pages = {690--697}
}

@article{r7,
  title = {Real-Time Neural {MPC}: Deep Learning Model Predictive Control for Quadrotors and Agile Robotic Platforms},
  author = {Salzmann, Tim and Kaufmann, Elia and Arrizabalaga, Jon and Pavone, Marco and Scaramuzza, Davide and Ryll, Markus},
  year = {2023},
  doi = {10.1109/LRA.2023.3246839},
  journal = {IEEE Robotics and Automation Letters},
  volume = {8},
  number = {4},
  pages = {2397--2404}
}

@inproceedings{r8,
  title = {Learning-based model predictive control on a quadrotor: Onboard implementation and experimental results},
  author = {Bouffard, Patrick and Aswani, Anil and Tomlin, Claire},
  year = {2012},
  doi = {10.1109/ICRA.2012.6225035},
  booktitle = {2012 IEEE International Conference on Robotics and Automation},
  pages = {279--284}
}

@article{r9,
  title = {Learning-Based Model Predictive Control: Toward Safe Learning in Control},
  author = {Hewing, Lukas and Wabersich, Kim P. and Menner, Marcel and Zeilinger, Melanie N.},
  year = {2020},
  doi = {10.1146/annurev-control-090419-075625},
  journal = {Annual Review of Control, Robotics, and Autonomous Systems},
  volume = {3},
  number = {1},
  pages = {269--296}
}

@inproceedings{r15,
  title = {Residual Reinforcement Learning for Robot Control},
  author = {Johannink, Tobias and Bahl, Shikhar and Nair, Ashvin and Luo, Jianlan and Kumar, Avinash and Loskyll, Matthias and Ojea, Juan Aparicio and Solowjow, Eugen and Levine, Sergey},
  year = {2019},
  doi = {10.1109/ICRA.2019.8794127},
  booktitle = {2019 International Conference on Robotics and Automation (ICRA)},
  pages = {6023--6029}
}

@inproceedings{r16,
  title = {Minimum snap trajectory generation and control for quadrotors},
  author = {Mellinger, Daniel and Kumar, Vijay},
  year = {2011},
  doi = {10.1109/ICRA.2011.5980409},
  booktitle = {2011 IEEE International Conference on Robotics and Automation},
  pages = {2520--2525}
}

@inproceedings{r17,
  title = {{NeuroBEM}: Hybrid Aerodynamic Quadrotor Model},
  author = {Bauersfeld, Leonard and Kaufmann, Elia and Föhn, Philipp and Sun, Sihao and Scaramuzza, Davide},
  year = {2021},
  doi = {10.15607/RSS.2021.XVII.042},
  booktitle = {Robotics: Science and Systems XVII}
}

@misc{r18,
  author = {Folk, Spencer and Paulos, James and Kumar, Vijay},
  title = {{RotorPy}: A {Python}-Based Multirotor Simulator with Aerodynamics for Education and Research},
  year = {2023},
  note = {arXiv:2306.04485; preprint},
  url = {https://arxiv.org/abs/2306.04485}
}

@article{r19,
  title = {Learning Model Predictive Control for Iterative Tasks. A Data-Driven Control Framework},
  author = {Rosolia, Ugo and Borrelli, Francesco},
  year = {2018},
  doi = {10.1109/TAC.2017.2753460},
  journal = {IEEE Transactions on Automatic Control},
  volume = {63},
  number = {7},
  pages = {1883--1896}
}

@article{r20,
  title = {Safe Reinforcement Learning Using Robust {MPC}},
  author = {Zanon, Mario and Gros, Sebastien},
  year = {2021},
  doi = {10.1109/TAC.2020.3024161},
  journal = {IEEE Transactions on Automatic Control},
  volume = {66},
  number = {8},
  pages = {3638--3652}
}

@inproceedings{r21,
  title = {Automated Tuning of Nonlinear Model Predictive Controller by Reinforcement Learning},
  author = {Mehndiratta, Mohit and Camci, Efe and Kayacan, Erdal},
  year = {2018},
  doi = {10.1109/IROS.2018.8594350},
  booktitle = {2018 IEEE/RSJ International Conference on Intelligent Robots and Systems (IROS)},
  pages = {3016--3021}
}

@book{r22,
  author = {Åström, Karl J. and Hägglund, Tore},
  title = {Advanced {PID} Control},
  publisher = {ISA},
  year = {2006},
  isbn = {978-1-55617-942-6},
  url = {https://portal.research.lu.se/en/publications/advanced-pid-control/}
}

@inproceedings{r23,
  title = {{PID} vs {LQ} control techniques applied to an indoor micro quadrotor},
  author = {Bouabdallah, S. and Noth, A. and Siegwart, R.},
  year = {2004},
  doi = {10.1109/IROS.2004.1389776},
  booktitle = {2004 IEEE/RSJ International Conference on Intelligent Robots and Systems (IROS) (IEEE Cat. No.04CH37566)},
  pages = {2451--2456 vol.3}
}

@article{r24,
  title = {{CasADi}: a software framework for nonlinear optimization and optimal control},
  author = {Andersson, Joel A. E. and Gillis, Joris and Horn, Greg and Rawlings, James B. and Diehl, Moritz},
  year = {2019},
  doi = {10.1007/s12532-018-0139-4},
  journal = {Mathematical Programming Computation},
  volume = {11},
  number = {1},
  pages = {1--36}
}

@article{r25,
  title = {{acados}—a modular open-source framework for fast embedded optimal control},
  author = {Verschueren, Robin and Frison, Gianluca and Kouzoupis, Dimitris and Frey, Jonathan and Duijkeren, Niels van and Zanelli, Andrea and Novoselnik, Branimir and Albin, Thivaharan and Quirynen, Rien and Diehl, Moritz},
  year = {2022},
  doi = {10.1007/s12532-021-00208-8},
  journal = {Mathematical Programming Computation},
  volume = {14},
  number = {1},
  pages = {147--183}
}

@article{r26,
  title = {A Real-Time Iteration Scheme for Nonlinear Optimization in Optimal Feedback Control},
  author = {Diehl, Moritz and Bock, Hans Georg and Schlöder, Johannes P.},
  year = {2005},
  doi = {10.1137/S0363012902400713},
  journal = {SIAM Journal on Control and Optimization},
  volume = {43},
  number = {5},
  pages = {1714--1736}
}

@article{r27,
  title = {{HPIPM}: a high-performance quadratic programming framework for model predictive control},
  author = {Frison, Gianluca and Diehl, Moritz},
  year = {2020},
  doi = {10.1016/j.ifacol.2020.12.073},
  journal = {IFAC-PapersOnLine},
  volume = {53},
  number = {2},
  pages = {6563--6569}
}

@inproceedings{r28,
  author = {Haarnoja, Tuomas and Zhou, Aurick and Abbeel, Pieter and Levine, Sergey},
  title = {Soft Actor-Critic: Off-Policy Maximum Entropy Deep Reinforcement Learning with a Stochastic Actor},
  booktitle = {Proceedings of the 35th International Conference on Machine Learning},
  volume = {80},
  pages = {1861--1870},
  year = {2018},
  url = {https://proceedings.mlr.press/v80/haarnoja18b.html}
}

@misc{r29,
  author = {{PX4 Developers}},
  title = {{PX4-SITL\_gazebo-classic}: Iris Model},
  year = {2026},
  note = {Source commit 807b67bb3007113ea52fd36ec2cefae4a5fa3f85; accessed 1 October 2026},
  url = {https://github.com/PX4/PX4-SITL_gazebo-classic/tree/807b67bb3007113ea52fd36ec2cefae4a5fa3f85}
}

@article{r30,
  title = {Nonlinear Model Predictive Control with Enhanced Actuator Model for Multi-Rotor Aerial Vehicles with Generic Designs},
  author = {Bicego, Davide and Mazzetto, Jacopo and Carli, Ruggero and Farina, Marcello and Franchi, Antonio},
  year = {2020},
  doi = {10.1007/s10846-020-01250-9},
  journal = {Journal of Intelligent \& Robotic Systems},
  volume = {100},
  number = {3-4},
  pages = {1213--1247}
}

@book{r31,
  author = {Åström, Karl J. and Murray, Richard M.},
  title = {Feedback Systems: An Introduction for Scientists and Engineers},
  publisher = {Princeton University Press},
  year = {2008},
  url = {https://www.cds.caltech.edu/~murray/FBS/First_Edition.html}
}

@misc{r32,
  author = {{acados Developers}},
  title = {{acados} {Python} Interface},
  year = {2026},
  note = {Official documentation; accessed 1 October 2026},
  url = {https://docs.acados.org/python_interface/index.html}
}

@misc{r33,
  author = {{acados Developers}},
  title = {{acados} Source Repository},
  year = {2026},
  note = {Commit 503364817c872d474ab5bed219c26760ac267769; accessed 1 October 2026},
  url = {https://github.com/acados/acados/commit/503364817c872d474ab5bed219c26760ac267769}
}

@misc{r35,
  author = {Haarnoja, Tuomas and Zhou, Aurick and Hartikainen, Kristian and Tucker, George and Ha, Sehoon and Tan, Jie and Kumar, Vikash and Zhu, Henry and Gupta, Abhishek and Abbeel, Pieter and Levine, Sergey},
  title = {Soft Actor-Critic Algorithms and Applications},
  year = {2019},
  note = {arXiv:1812.05905v2; preprint},
  url = {https://arxiv.org/abs/1812.05905v2}
}

@article{r36,
  author = {Raffin, Antonin and Hill, Ashley and Gleave, Adam and Kanervisto, Anssi and Ernestus, Maximilian and Dormann, Noah},
  title = {{Stable-Baselines3}: Reliable Reinforcement Learning Implementations},
  journal = {Journal of Machine Learning Research},
  volume = {22},
  number = {268},
  pages = {1--8},
  year = {2021},
  url = {https://jmlr.org/papers/volume22/20-1364/20-1364.pdf}
}

@inproceedings{r37,
  author = {Agarwal, Rishabh and Schwarzer, Max and Castro, Pablo Samuel and Courville, Aaron and Bellemare, Marc G.},
  title = {Deep Reinforcement Learning at the Edge of the Statistical Precipice},
  booktitle = {Advances in Neural Information Processing Systems},
  volume = {34},
  pages = {29304--29320},
  year = {2021},
  url = {https://papers.nips.cc/paper/2021/hash/f514cec81cb148559cf475e7426eed5e-Abstract.html}
}

@misc{r38,
  author = {{Stable-Baselines3 Developers}},
  title = {{SAC}: Version 2.7.0 Documentation and Source},
  year = {2026},
  note = {Accessed 1 October 2026},
  url = {https://stable-baselines3.readthedocs.io/en/v2.7.0/modules/sac.html}
}

@article{r39,
  title = {Performance-Oriented Model Learning for Data-Driven {MPC} Design},
  author = {Piga, Dario and Forgione, Marco and Formentin, Simone and Bemporad, Alberto},
  year = {2019},
  doi = {10.1109/LCSYS.2019.2913347},
  journal = {IEEE Control Systems Letters},
  volume = {3},
  number = {3},
  pages = {577--582}
}

@article{r40,
  title = {Deep Reinforcement Learning That Matters},
  author = {Henderson, Peter and Islam, Riashat and Bachman, Philip and Pineau, Joelle and Precup, Doina and Meger, David},
  year = {2018},
  doi = {10.1609/aaai.v32i1.11694},
  journal = {Proceedings of the AAAI Conference on Artificial Intelligence},
  volume = {32},
  number = {1}
}

@article{r41,
  title = {Disturbance Observer Based Control for Nonlinear Systems},
  author = {Chen, W.-H.},
  year = {2004},
  doi = {10.1109/TMECH.2004.839034},
  journal = {IEEE/ASME Transactions on Mechatronics},
  volume = {9},
  number = {4},
  pages = {706--710}
}

@inproceedings{r42,
  title = {Overcoming Exploration in Reinforcement Learning with Demonstrations},
  author = {Nair, Ashvin and McGrew, Bob and Andrychowicz, Marcin and Zaremba, Wojciech and Abbeel, Pieter},
  year = {2018},
  doi = {10.1109/ICRA.2018.8463162},
  booktitle = {2018 IEEE International Conference on Robotics and Automation (ICRA)},
  pages = {6292--6299}
}

@inproceedings{r43,
  author = {Yuan, Xiu and Mu, Tongzhou and Tao, Stone and Fang, Yunhao and Zhang, Mengke and Su, Hao},
  title = {Policy Decorator: Model-Agnostic Online Refinement for Large Policy Model},
  booktitle = {International Conference on Learning Representations},
  year = {2025},
  url = {https://proceedings.iclr.cc/paper_files/paper/2025/hash/45c361d4117d598d4bb6568b407e9ac9-Abstract-Conference.html}
}

@misc{r44,
  author = {Ma, Guozheng and Li, Lu and Wang, Haoyu and Liu, Zixuan and Bacon, Pierre-Luc and Tao, Dacheng},
  title = {What Makes Value Learning Efficient in Residual Reinforcement Learning?},
  year = {2026},
  note = {arXiv:2602.10539v1; preprint},
  url = {https://arxiv.org/abs/2602.10539}
}

@inproceedings{r45,
  author = {Fujimoto, Scott and van Hoof, Herke and Meger, David},
  title = {Addressing Function Approximation Error in Actor-Critic Methods},
  booktitle = {Proceedings of the 35th International Conference on Machine Learning},
  volume = {80},
  pages = {1587--1596},
  year = {2018},
  url = {https://proceedings.mlr.press/v80/fujimoto18a.html}
}

@article{r46,
  author = {Nicoud, F. and Ducros, F.},
  title = {Subgrid-Scale Stress Modelling Based on the Square of the Velocity Gradient Tensor},
  journal = {Flow, Turbulence and Combustion},
  year = {1999},
  volume = {62},
  number = {3},
  pages = {183--200},
  doi = {10.1023/A:1009995426001},
  url = {https://link.springer.com/article/10.1023/A:1009995426001}
}

@misc{r47,
  author = {{OpenCFD Developers}},
  title = {{OpenFOAM} v2512},
  year = {2025},
  url = {https://www.openfoam.com}
}

@misc{r48,
  author = {{ParaView Developers}},
  title = {{ParaView} 6.1 User Guide: Filtering Data and Data Display},
  year = {2026},
  note = {Versioned documentation; accessed 4 October 2026},
  url = {https://docs.paraview.org/en/v6.1.0/UsersGuide/filteringData.html},
  howpublished = {Official user guide and reference manual. Animation: \url{https://docs.paraview.org/en/v6.1.0/UsersGuide/animation.html}; color mapping: \url{https://docs.paraview.org/en/v6.1.0/ReferenceManual/colorMapping.html}}
}

@article{r49,
  author = {Crameri, Fabio and Shephard, Grace E. and Heron, Philip J.},
  title = {The Misuse of Colour in Science Communication},
  journal = {Nature Communications},
  year = {2020},
  volume = {11},
  pages = {5444},
  doi = {10.1038/s41467-020-19160-7},
  url = {https://www.nature.com/articles/s41467-020-19160-7}
}

@article{r50,
  author = {Nathan, J{\"o}rn and Masson, Christian and Dufresne, Louis},
  title = {Near-Wake Analysis of Actuator Line Method Immersed in Turbulent Flow Using Large-Eddy Simulations},
  journal = {Wind Energy Science},
  year = {2018},
  volume = {3},
  number = {2},
  pages = {905--917},
  doi = {10.5194/wes-3-905-2018},
  url = {https://wes.copernicus.org/articles/3/905/2018/}
}

\clearpage
\setcounter{page}{1}
\renewcommand{\thepage}{S-\arabic{page}}
\setcounter{section}{0}
\setcounter{subsection}{0}
\setcounter{equation}{0}
\setcounter{figure}{0}
\setcounter{table}{0}
\renewcommand{\thesection}{S\arabic{section}}
\renewcommand{\theequation}{S\arabic{equation}}
\renewcommand{\thefigure}{S\arabic{figure}}
\renewcommand{\thetable}{S\arabic{table}}
\renewcommand{\theHsection}{supp.\arabic{section}}
\renewcommand{\theHequation}{supp.\arabic{equation}}
\renewcommand{\theHfigure}{supp.\arabic{figure}}
\renewcommand{\theHtable}{supp.\arabic{table}}

\section{Extended related work and baseline implementation}\label{sec:extended_related}

\subsection*{Aerodynamics and learned prediction}

Drag-aware flatness \cite{r1}, adaptive NMPC \cite{r6}, data-driven MPC \cite{r3}, and neural MPC \cite{r7} address different parts of the model--control chain. Onboard learning-based MPC provides an implementation precedent \cite{r8}; NeuroBEM combines physics and learned aerodynamic residuals \cite{r17}. These results motivate residual augmentation, but do not supply the outcome of our frozen comparison. Control-oriented identification distinguishes prediction fit from performance under feedback \cite{r39}; both must be measured here.

Drag-aware flatness and minimum-snap trajectory design motivate consistent references \cite{r1,r16}, rather than exact inversion of the present aerodynamic plant. Disturbance-aware MPC \cite{r2} and adaptive NMPC \cite{r6} differ from our frozen residual interface. Safe learning MPC \cite{r9,r20} requires assumptions and safeguards not established by a successful solve here; iterative-task learning MPC \cite{r19} is also a different learning mechanism. Hybrid aerodynamic modeling \cite{r17} and the RotorPy simulator preprint \cite{r18} motivate separating modeled rotor loads from injected environment loads.

\subsection*{Residual policy and fixed compensation}

Residual RL augments a baseline command \cite{r15}; other MPC--RL combinations constrain learning or tune controller parameters \cite{r20,r21}. SAC uses maximum-entropy off-policy learning \cite{r28,r35}, with a versioned implementation in Stable-Baselines3 \cite{r36,r38}. Its benchmark success does not guarantee an increment over a strong local baseline. Few-run evaluation warrants separate seed reporting \cite{r37,r40}. Disturbance-observer control motivates a causal compensation channel \cite{r41}, but an acceleration innovation is not automatically an identified external force.

Demonstration-based initialization \cite{r42} and frozen-policy refinement \cite{r43} motivate preserving an existing controller during learning. The residual-value-learning preprint \cite{r44} proposes specific cold-start and scale-mismatch explanations in other robotic tasks. We use it to formulate a possible explanation, not to diagnose our critic or claim implementation of that complete algorithm. Value-approximation error \cite{r45} and the residual-value-learning preprint \cite{r44} supply competing hypotheses, not measured causes here.

\subsection{PID details}\label{sec:pid_details}

\begin{equation}\label{eq:pid_force}\mathbf F_{\rm cmd}^{W}=m(\mathbf a_{\rm PID}+g\mathbf e_3),\qquad T_{\rm cmd}=\|\mathbf F_{\rm cmd}^{W}\|_2.\end{equation}

\begin{equation}\label{eq:pid_moment}\mathbf M_{\rm cmd}^{B}=K_R\operatorname{Log}(R^TR_d)^\vee-K_\omega\boldsymbol\omega.\end{equation}

\begin{equation}\label{eq:pid_gains}\begin{aligned}K_p&=\omega_o^2+2\zeta\omega_op_i,\quad K_i=p_i\omega_o^2,\\K_v&=2\zeta\omega_o+p_i,\\K_{R,j}&=I_{jj}\omega_r^2,\quad K_{\omega,j}=2\zeta I_{jj}\omega_r.\end{aligned}\end{equation}

The gain relations apply componentwise and match $(s^2+2\zeta\omega_os+\omega_o^2)(s+p_i)$, where $s$ is the Laplace variable, $\omega_o,\omega_r$ are response rates, $\zeta$ damping, and $p_i$ the integral-pole rate. $\operatorname{Log}$ is the rotation logarithm; vee converts its skew matrix to a three-vector. $R_d$ uses desired force direction and zero yaw. This candidate construction excludes lag and nonlinear coupling \cite{r22,r31}. Integral updates stop during allocation saturation; otherwise each component is clipped to $\pm0.277778$ m\,$\cdot$\,s. PID-only development compares $\omega_r=8,12,16$ rad/s on 12 zero-force runs and checks eight signed-force hover cases. The selected 16 rad/s is best within that finite development set, not globally optimal.

\noindent\begin{minipage}{\columnwidth}
\captionof{table}{Frozen PID gains; units follow the requested acceleration and moment.}\label{tab:pid_gains}\begin{center}\footnotesize\setlength{\tabcolsep}{3pt}\begin{tabular}{lrrr}\toprule
Gain & X / roll & Y / pitch & Z / yaw\\\midrule
$K_p$ (s$^{-2}$) & 4.8 & 4.8 & 10.2\\
$K_i$ (s$^{-3}$) & 0.8 & 0.8 & 1.8\\
$K_v$ (s$^{-1}$) & 4.2 & 4.2 & 6.2\\
$K_R$ (N m/rad) & 7.6985054 & 7.8468966 & 14.745639\\
$K_\omega$ (N m\,$\cdot$\,s/rad) & 0.96231317 & 0.98086207 & 1.8432049\\\bottomrule\end{tabular}\end{center}
\end{minipage}

Historical PID trajectories were reused after a 220-case audit of scenario identity, plant and asset hashes, current performance metrics, and current success criteria. The controller was not rerun. This supports case-matched performance comparison with current nominal NMPC, but does not establish equivalence between historical and current nominal implementations. Historical PID timing remains a separate acquisition.

The original PID--historical nominal batch contains 220 pairs. Its mean-IAE reductions are 28.31\%, 25.70\%, 53.69\%, and 58.66\% for H/L/C/F8. H/L peaks nevertheless increase slightly. Smaller late-window error explains the integrated advantage arithmetically, while moving-task gains also occur without force. Preview and controller structure change together: PID receives current-reference feedback, NMPC a future window. The comparison cannot assign the entire gain to disturbance rejection or optimization alone. Figure~\ref{fig:fig02_absolute_grid} uses the current nominal identity, whose means must not be substituted into those historical percentages.

\section{Optimization, observations, and reward}\label{sec:rl_details}

\begin{equation}\label{eq:augmented}\mathbf z=\operatorname{col}(\mathbf x,\mathbf h)\in\mathbb R^{21},\qquad\mathbf h_{i+1|k}=\eta(\mathbf u_{i|k}).\end{equation}

\begin{equation}\label{eq:argmin}U_k^*=\operatorname*{arg\,min}_{U_k}J_k,\quad U_k=\operatorname{col}(\mathbf u_{0|k},\ldots,\mathbf u_{N-1|k}).\end{equation}

\begin{equation}\label{eq:constraints}\begin{aligned}\mathbf x_{i+1|k}&=F_\Delta(\mathbf x_{i|k},\mathbf u_{i|k};\mathbf d_{i|k}),\\0&\le u_{j,i|k},\Omega_{j,i|k}\le\Omega_{\max}.\end{aligned}\end{equation}

Commands are bounded for $i=0,\ldots,N-1$, actual speeds at $i=0,\ldots,N$. The 17-component reference has blocks $3+3+4+3+4$: position, velocity, quaternion, angular rate, and thrust--moment reference; its final block is not motor state. The positive diagonal $4\times4$ matrix $D_u$ converts thrust/moment errors to dimensionless costs. Each diagonal entry is the maximum absolute deviation of that output from hover over the 16 vertices of the allowable motor-thrust box. This is an actuator-authority scale derived from the fixed allocation, not a fitted performance parameter. The optimization residual is

\begin{equation}\label{eq:tracking_residual}\begin{aligned}\mathbf e_x=\operatorname{col}\big(& (\mathbf p-\mathbf p_{\rm ref})/s_p,\\& (\mathbf v-\mathbf v_{\rm ref})/s_v,\\& 2\operatorname{vec}(\mathbf q_e)/s_q,\\& (\boldsymbol\omega-\boldsymbol\omega_{\rm ref})/s_\omega\big)\in\mathbb R^{12}.\end{aligned}\end{equation}

$\mathbf q_e=\mathbf q_{\rm ref}^{-1}\otimes\mathbf q$. Scales $(s_p,s_v,s_q,s_\omega)=(0.141421,0.707107,0.246827,1.414214)$ use m, m/s, attitude coordinates, and rad/s; $w_u,w_{\Delta u},w_N=0.04,0.05,5$. These are frozen relative-priority choices. The current 29-component stage parameter adds supervised and actor six-vectors to the reference. One SQP-RTI update uses the explicit cost Hessian $2J_r^TJ_r$, where $J_r$ is the least-squares-residual Jacobian; it is unrelated to $J_k$ or the network \cite{r24,r25,r26,r27}. Status and fallback are separate from tracking and optimality.

\begin{equation}\label{eq:observation}\begin{aligned}\mathbf o_k=\operatorname{col}\big(&\mathbf e_{p,k},\mathbf e_{v,k},\mathbf q_{e,k}^{c},\mathbf e_{\omega,k},\\&\mathbf u_{k-1}^{\rm exec}/\bar\Omega_h-\mathbf1_4,\mathbf a_{k-1},\mathbf v_{{\rm ref},k}\big).\end{aligned}\end{equation}

The dimensions are $3+3+4+3+4+m_a+3$, with $m_a=4$ external or 6 internal. In Eq.~\eqref{eq:observation}, $\mathbf e_p=\mathbf p-\mathbf p_{\rm ref}$, $\mathbf e_v=\mathbf v-\mathbf v_{\rm ref}$, and $\mathbf e_\omega=\boldsymbol\omega-\boldsymbol\omega_{\rm ref}$; the canonical quaternion $\mathbf q_e^c$ chooses a nonnegative scalar part, with the largest-magnitude vector component setting the sign when $|q_{e,w}|\le10^{-12}$. $\bar\Omega_h$ is mean hover speed; $\mathbf1_4$ is a four-vector of ones. Reset and fallback clear action history. The output and history conventions are preserved under structural masks. Position, velocity, angular-rate differences and reference velocity are unscaled in this original input; physical units remain. This is the sealed implementation, not the scaled NMPC cost residual.

\begin{equation}\label{eq:attitude_metric}e_R(t)=2\arccos\operatorname{clip}\big(|\tilde{\mathbf q}(t)^T\tilde{\mathbf q}_{\rm ref}(t)|,0,1\big).\end{equation}

For reported attitude error, $\tilde{\mathbf q}=\mathbf q/\|\mathbf q\|_2$ normalizes stored quaternions; Eq.~\eqref{eq:attitude_metric} is the sign-invariant principal rotation angle in radians. Its integrated RMS is different from the quaternion-vector cost/reward. Velocity and angular-rate RMSE use the same 0--8 s integrated RMS of $\|\mathbf v-\mathbf v_{\rm ref}\|_2$ and $\|\boldsymbol\omega-\boldsymbol\omega_{\rm ref}\|_2$; the latter is the stored body-coordinate difference convention, without an additional reference-to-current-body rotation.

\begin{equation}\label{eq:reward}r_k=-\sum_{j=1}^{11}w_j c_{j,k}.\end{equation}

Costs in order are squared position/velocity/quaternion-vector/angular-rate errors; squared normalized command and command increment; squared normalized action and action increment; clipped-motor count; solve-failure/fallback indicator; termination indicator. Quaternion-vector reward omits the NMPC factor 2. Weights are $1,0.1,0.1,0.01,0.001,0.01,0.001,0.001,0.1,10,10$; position, velocity, and angular-rate coefficients absorb inverse-squared physical units, while the quaternion-vector cost is dimensionless. Costs are not multiplied by $\Delta$. Error above 5 m aborts; otherwise 400 steps truncate. These objective/cutoff choices do not certify safety.

\section{Supervised fit, assessment, and training accounting}\label{sec:prediction_details}

\begin{figure*}[!t]
\begin{subfigure}[t]{.32\textwidth}
  \centering
  \includegraphics[width=\linewidth]{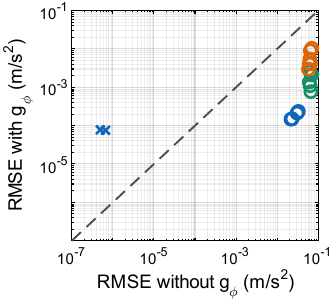}
  \caption{Body $x$ linear residual}\label{fig:figS01_prediction_panel1}
\end{subfigure}\hfill\begin{subfigure}[t]{.32\textwidth}
  \centering
  \includegraphics[width=\linewidth]{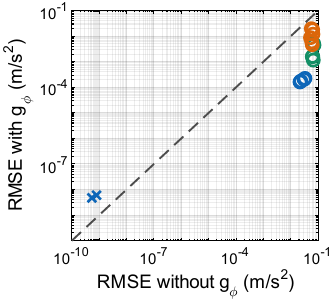}
  \caption{Body $y$ linear residual}\label{fig:figS01_prediction_panel2}
\end{subfigure}\hfill\begin{subfigure}[t]{.32\textwidth}
  \centering
  \includegraphics[width=\linewidth]{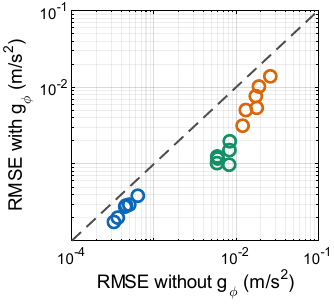}
  \caption{Body $z$ linear residual}\label{fig:figS01_prediction_panel3}
\end{subfigure}\par\smallskip
\begin{subfigure}[t]{.32\textwidth}
  \centering
  \includegraphics[width=\linewidth]{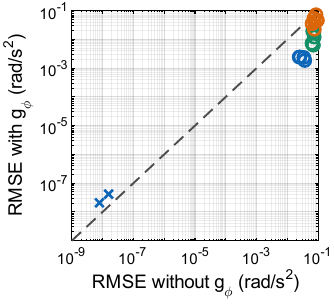}
  \caption{Body $x$ angular residual}\label{fig:figS01_prediction_panel4}
\end{subfigure}\hfill\begin{subfigure}[t]{.32\textwidth}
  \centering
  \includegraphics[width=\linewidth]{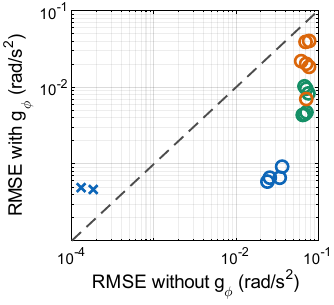}
  \caption{Body $y$ angular residual}\label{fig:figS01_prediction_panel5}
\end{subfigure}\hfill\begin{subfigure}[t]{.32\textwidth}
  \centering
  \includegraphics[width=\linewidth]{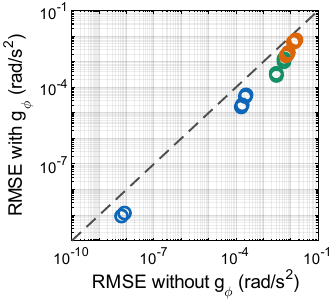}
  \caption{Body $z$ angular residual}\label{fig:figS01_prediction_panel6}
\end{subfigure}
{\centering
\predictionlegend
}
\caption{Learned and zero-residual absolute RMSE on all 18 groups for each physical component. Equality is a reference, not a controller difference. Components 1--3 are body linear $x$, $y$, $z$ residuals (m/s$^2$); 4--6 are angular residuals (rad/s$^2$). Task colors and pass/fail markers match Figure~\ref{fig:prediction_main}. Each panel contains all 18 records. Points below the diagonal have lower learned error; those above it fail the no-margin criterion.}\label{fig:figS01_prediction}
\end{figure*}

Figure~\ref{fig:figS01_prediction} extends the main prediction plot to all six physical outputs. It tests componentwise prediction error, not tracking: 100 of 108 checks pass, and the eight failed records are retained in Table~\ref{tab:prediction_fail}. Panels (a), (b), (d) and (e) each contain two L points above the diagonal; panels (c) and (f) improve or match all 18 groups. Thus prediction gains are widespread but do not satisfy the all-component criterion. Very small failed errors remain failures of the fixed no-margin rule; overlap on a logarithmic axis does not remove a record.

An independently constructed nominal NMPC, identified in the collection manifest as \texttt{C17\_\allowbreak ATTITUDE\_\allowbreak VECTOR\_\allowbreak M0\_\allowbreak DEVELOPMENT\_\allowbreak ONLY}, supplies the still-air trajectories. It uses the corrected quaternion-vector cost and a fresh solver for each group; it is neither PID nor the residual controller. At each 20 ms collection instant, plant and nominal derivatives are evaluated at the same 17-state and executed command. Their linear acceleration difference is rotated into body axes and concatenated with the body angular-acceleration difference. Thus the target is an instantaneous dynamics discrepancy, not a velocity finite difference. Training L, C and F8 at multiplier 0.75 in X and Y gives six groups of 400. Validation and sealed partitions use multiplier 0.875 in $45^\circ$ and X, three groups each. Final assessment uses L, C and F8, directions X, Y and $45^\circ$, and multipliers 13/\allowbreak 16 and 15/\allowbreak 16: 18 groups of 400. Multipliers scale references, not force. Each group resets state, hover motors, and solver. Distinct reference tables do not imply disjoint states or IID samples.

For training rows $n=1,\ldots,N_{\rm tr}$, let $\bar{\boldsymbol\chi}_n=(\boldsymbol\chi_n-\boldsymbol\mu_\chi)\oslash\boldsymbol\sigma_\chi$ and $\bar{\mathbf y}_n=(\mathbf y_n-\boldsymbol\mu_y)\oslash\boldsymbol\sigma_y$. The dimensionless fitting objective is
\begin{equation}\label{eq:fit_loss}
\mathcal L_{\rm fit}(\phi)=\frac{1}{6N_{\rm tr}}\sum_{n=1}^{N_{\rm tr}}\|G_\phi(\bar{\boldsymbol\chi}_n)-\bar{\mathbf y}_n\|_2^2,
\quad N_{\rm tr}=2400.
\end{equation}
Normalization uses componentwise population standard deviations (divisor $N_{\rm tr}$); an exactly zero deviation is replaced by one. No validation or sealed-assessment row enters these statistics or the fit. L-BFGS-B uses the analytic gradient of this full-batch loss. Mirror averaging in Eq.~\eqref{eq:mirror} is applied after inverse normalization at deployment; it is not an extra term in the fitting loss.

For each supervised case, the actuator estimate starts at the hover rotor speeds, the previous executed command is hover, and stored state/command predictions are cleared. The first 21-node guess starts at the current augmented state and advances the corrected RK4 map for 20 intervals with that command and the single current residual held fixed (zero actor correction for supervision). Stagewise sampling then uses node 0 and nodes 1--19; node 20 copies node 19's residual. After a successful solve, both predictions shift by one node, their final entries are repeated, and the first state is overwritten. This distinguishes initialization from subsequent warm starting without using future plant states.

Seed 20260913 initializes zero biases and uniform layer weights within $\pm\sqrt{6/\allowbreak 42}$ and $\pm\sqrt{6/\allowbreak 38}$. Means/standard deviations and fitting use only 2400 training rows. Limited-memory BFGS with bounds (L-BFGS-B) caps iterations/evaluations at 15000 iterations and 300001 evaluations, with ftol $2.220446049250313\times10^{-9}$, gtol $10^{-5}$, maxcor 10, maxls 20. It stops at 5994 iterations and 6201 evaluations by relative loss reduction, normalized loss 0.00024738119618597184. The iteration and evaluation caps bound fitting cost; ftol and gtol are numerical stopping rules, not predictor-qualification thresholds. The terminal fit is retained without checkpoint selection; width 32 is a project budget choice. Residual-model precedents motivate the method, not this exact width \cite{r3,r7}.

\noindent\begin{minipage}{\columnwidth}
\captionof{table}{All eight failed group--component checks. Components 1--3: m/s$^2$; 4--6: rad/s$^2$.}\label{tab:prediction_fail}\begin{center}\footnotesize\setlength{\tabcolsep}{3pt}\begin{tabular}{lrrr}\toprule
Group & Comp. & Learned & Zero\\\midrule
L\_final\_13of16\_X & 2 & 6.006595e-09 & 7.967948e-10\\
L\_final\_13of16\_X & 4 & 4.133982e-08 & 1.560322e-08\\
L\_final\_13of16\_Y & 1 & 7.775955e-05 & 4.924555e-07\\
L\_final\_13of16\_Y & 5 & 4.902984e-04 & 1.295377e-04\\
L\_final\_15of16\_X & 2 & 4.726487e-09 & 5.394416e-10\\
L\_final\_15of16\_X & 4 & 2.067118e-08 & 7.858384e-09\\
L\_final\_15of16\_Y & 1 & 7.558663e-05 & 6.632934e-07\\
L\_final\_15of16\_Y & 5 & 4.652327e-04 & 1.820623e-04\\\bottomrule\end{tabular}\end{center}
\end{minipage}

The three retained Seed 1 actors (external, internal and retrained without supervision) each complete 6400 interactions, 6368 updates and 16 episodes. Internal stagewise and fixed-removal evaluation reuse the internal-cyclewise weights. Training repeats four frozen cases four times, not 16 sampled environments. Table~\ref{tab:training_cases} lists the four training cases.
\noindent\begin{minipage}{\columnwidth}\captionof{table}{Frozen training cases shared by the retained Seed 1 policies.}\label{tab:training_cases}\begin{center}\footnotesize\setlength{\tabcolsep}{3pt}\begin{tabular}{lrrrr}\toprule
 & H & L & C & F8\\\midrule
Force/weight & 0.05253343 & 0.04267916 & 0.06111799 & 0.05832588\\
Onset (s) & 1.320 & 0.835 & 1.235 & 1.065\\
Duration (s) & 0.6 & 0.6 & 0.6 & 0.6\\\bottomrule
\end{tabular}\end{center}\end{minipage}
Training/testing load coverage differs; it is not an isolated cause of policy performance.

The first and last complete four-task cycles compare the same training case bank. Logged reward accounting excludes entropy and does not measure critic gradients. Improved return can be dominated by action regularization while position costs move in different directions. The data therefore explain a metric divergence without proving that those weights caused the deployed policy loss.

\section{All retained identities on the 220-condition grid}\label{sec:all_grid}

\begin{table*}[!t]
\caption{Case-mean absolute errors, 220 cases per identity. All policy rows use Seed 1 (6201). Bold is the smallest unrounded value in each displayed column; PID metrics are recomputed from archived trajectories using the current integral definitions.}\label{tab:all_grid}\begin{center}\footnotesize\setlength{\tabcolsep}{3pt}\begin{tabular}{lrrrr}\toprule
Method & IAE (m\,$\cdot$\,s) & RMSE (m) & Peak (m) & End (mm)\\\midrule
Nominal NMPC & 0.216494 & 0.056253 & 0.178509 & \textbf{0.020542}\\
Supervised cycle & 0.207016 & 0.056876 & 0.183818 & 0.038183\\
Supervised stage & \textbf{0.199679} & 0.055245 & 0.178996 & 0.026468\\
External RL Seed 1 & 0.207266 & 0.056862 & 0.183774 & 0.130495\\

Internal cycle Seed 1 & 0.220089 & 0.056858 & 0.183045 & 3.177928\\

Internal stage Seed 1 & 0.212478 & \textbf{0.055232} & 0.178265 & 3.017150\\

Fixed removal Seed 1 & 0.229743 & 0.056360 & 0.177949 & 3.167052\\

Retrained Seed 1 & 0.234491 & 0.056384 & \textbf{0.177719} & 4.015153\\

PID & 0.404989 & 0.077588 & 0.214010 & 19.860162\\\bottomrule\end{tabular}\end{center}
\end{table*}

The table retains unfavorable averages and secondary measures. RMSE/peak improvements do not reverse a higher IAE. The nine displayed identities comprise 1980 method--case records on one 220-case bank. Derived pairings reuse these runs and are not additional experiments. Other seeds remain in the engineering archive. The historical PID/current-nominal bridge is a case-matched trajectory-reuse audit, not a fresh common-timing batch. Full scenario rows and reference/error arrays remain in the engineering evidence package.

\subsection{Structural interventions}\label{sec:structure}

V1 masks only the actor's three angular outputs while retaining all six action-history components. V2 uses $\rho=0.95$ to attenuate only the actor correction along the prediction horizon; the supervised residual is unchanged. Both reuse the original actor weights. The frozen decay factor tests reduced persistence of actor injection, not a sensitivity-optimized value.

\noindent\begin{minipage}{\columnwidth}
\captionof{table}{Matched 220-case structural identities. Counts are lower/higher IAE against parent and cyclewise supervised control.}\label{tab:structural}\begin{center}\footnotesize\setlength{\tabcolsep}{3pt}\begin{tabular}{lrrr}\toprule
Variant & IAE (m\,$\cdot$\,s) & vs parent & vs supervised\\\midrule
V1 & 0.220085 & 150, 70 & 0, 220\\

V2 & \textbf{0.216173} & 220, 0 & 0, 220\\

\bottomrule\end{tabular}\end{center}
\end{minipage}

\begin{equation}\label{eq:structural}\begin{aligned}P_i^{\rm original}&=I_6,\quad P_i^{V1}=\operatorname{diag}(1,1,1,0,0,0),\\P_i^{V2}&=\rho^iI_6.\end{aligned}\end{equation}

For Seed 1, V2 improves all 220 parent-policy cases, but none against cyclewise supervision. V1 improves 150 parent cases and worsens 70, with almost unchanged mean IAE; it also improves none against supervision. Thus attenuating actor persistence helps this policy without supplying a net benefit over the matched non-RL control. Fixed removal and retraining remain separate rows in Table~\ref{tab:all_grid}: changing a deployed baseline and relearning a policy answer different questions.

\section{Fixed compensation and separate confirmation}\label{sec:fixed_confirmation}

At reset, innovation and filtered estimate are zero until a preceding executed transition exists. The 24-condition development bank (D24) compares $\beta=0.2$ against 1 and $\kappa=0.5,1,1.5$. For $\beta=0.2$, the filter time constant is $-0.02/\ln0.8=0.089628$ s; this is an algebraic interpretation, not identified external-load dynamics. Gain precedes saturation, with $\mathbf s_v=(0.572,0.546,0.053)$ m/s$^2$ and $\rho=0.95$. Gain 1.5 is selected before T32. Nominal, low-/high-force development cases and gain-pulse interventions inform interpretation, not independent confirmation of RL.

The four absolute means and matched metric minima are reported once in Table~\ref{tab:t32_main}.

T32 seed 53001 freezes eight scenes per trajectory: three low, three medium, and two high force bands $[0.02,0.08]$, $[0.08,0.16]$, $[0.16,0.24]$. Onset is 0.75--2.75 s and duration 0.4--1.2 s. Exact requested/effective times remain in the scenario table. It is a separate confirmation for the selected fixed compensator. The frozen gates require 32 successes, at least 3\% mean IAE gain versus stagewise supervision, positive gain versus cyclewise, at least 24 nonworse cases (tolerance $10^{-9}$ m\,$\cdot$\,s), nominal/tail error limits 0.2 mm, peak mean at most 1.02 times stagewise, and a negative upper descriptive resampling bound. The descriptive interval resamples within trajectory--intensity-band strata (10000 repetitions, seed 53002), retaining the paired methods and reporting the 2.5/\allowbreak 97.5 percentiles. It summarizes the fixed grid, not an unrestricted population. These engineering gates were not statistical-significance thresholds.

\begin{table}[!t]\centering\small
\caption{T32 paired mean IAE differences and 95\% stratified descriptive resampling intervals (m\,$\cdot$\,s). Each row subtracts the named comparator from fixed gain 1.5 on the same 32 cases; negative values favor gain 1.5.}\label{tab:t32_intervals}
\setlength{\tabcolsep}{3pt}
\begin{tabular}{lrrr}\toprule
Comparator & Mean & Lower & Upper\\\midrule
Supervised stage & $-0.045698$ & $-0.049957$ & $-0.041606$\\
Supervised cycle & $-0.051934$ & $-0.056577$ & $-0.047453$\\
Fixed gain 1 & $-0.002981$ & $-0.003692$ & $-0.002322$\\\bottomrule
\end{tabular}\end{table}
Table~\ref{tab:t32_intervals} reports the actual statistic used by the frozen gate, the mean paired IAE difference. Resampling preserves the 12 trajectory--intensity strata and each stratum's size (three low, three medium, two high cases per task); its interval endpoints are the 2.5th and 97.5th percentiles of 10000 resampled means. This describes variation within the prescribed bank. It does not treat time steps or method pairings as independent trials, and it does not turn the all-case position benefit into an all-state superiority claim.

Gain 1.5 raises angular-rate RMSE about 10.19\% versus stagewise supervision and slightly raises mean attitude error. Hence tracking improvement is not uniform state improvement. Gain-pulse changes in two windows $[2.20,2.40)$ and $[3.10,3.30)$ s probe the same compensator locally, not all possible gains or times. The original T32 secondary force/recovery analysis inherited an incorrect common window; the erratum uses each actual scene event and remains a post hoc supplementary analysis. The original gates are not retroactively changed. A missing pre-run numerical-file enumeration was supplemented by matching an earlier historical receipt. Baseline and observer timing boundaries differ, so they do not support overhead ranking.

\subsection{Existing-state and command activity check}\label{sec:activity}

The 128 T32 raw logs were read without rerunning dynamics. Position IAE, attitude RMS and angular-rate RMS were reintegrated from 1601 states and references per run and agree with the retained values to $10^{-9}$ in their respective units. All 400 command samples were retained. Let $\mathbf u_k$ be the executed four-motor speed command. The added descriptive activity diagnostic is
\begin{equation}\label{eq:command_activity}
A_u=\sqrt{\frac1{399}\sum_{k=1}^{399}\|\mathbf u_k-\mathbf u_{k-1}\|_2^2}.
\end{equation}
This is a command-increment RMS in rad/s per sample change, not angular acceleration or a frequency-selective jitter measure. No filtering or detrending is applied. Case means are 6.726359 (stagewise supervision), 9.154823 (gain 1) and 10.748675 rad/s (gain 1.5). Gain 1.5 exceeds stagewise in 32 of 32 scenes. The same-case mean over actual post-force windows is 4.416062 versus 5.227777 rad/s. Thus the larger commands are not solely an average artifact from imposing the common grid's 2--3 s window.

\begin{figure*}[!t]\centering
\begin{subfigure}[t]{.48\textwidth}
  \centering
  \includegraphics[width=\linewidth]{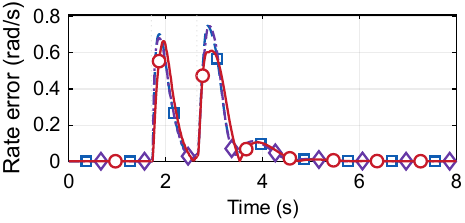}
  \caption{H: angular-rate error}\label{fig:figS03_fixed_activity_panel1}
\end{subfigure}\hfill\begin{subfigure}[t]{.48\textwidth}
  \centering
  \includegraphics[width=\linewidth]{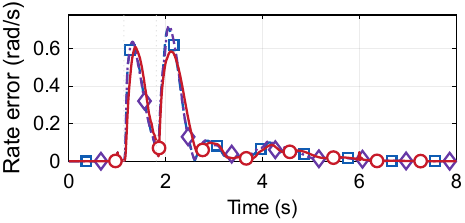}
  \caption{F8: angular-rate error}\label{fig:figS03_fixed_activity_panel2}
\end{subfigure}\par\smallskip
\begin{subfigure}[t]{.48\textwidth}
  \centering
  \includegraphics[width=\linewidth]{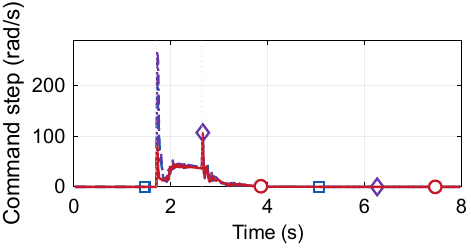}
  \caption{H: motor-command increment}\label{fig:figS03_fixed_activity_panel3}
\end{subfigure}\hfill\begin{subfigure}[t]{.48\textwidth}
  \centering
  \includegraphics[width=\linewidth]{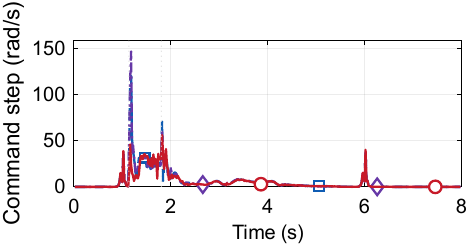}
  \caption{F8: motor-command increment}\label{fig:figS03_fixed_activity_panel4}
\end{subfigure}\par\smallskip{\small\curvekey{nomcolor}{dashed}{$\square$}{Fixed gain 1}\curvekey{fixedgaincolor}{dash dot}{$\diamond$}{Fixed gain 1.5}\curvekey{stagecolor}{solid}{$\circ$}{Supervised stagewise}}

\caption{Raw reference-relative angular-rate error norm and consecutive motor-command increment norm for the highest-force H and F8 confirmation scenes, selected by scenario metadata with lexicographic tie-breaking before examining performance. Dotted lines use each scene's actual onset/removal. Full 0--8 s records remain unsmoothed; panel scales differ. The compensator produces larger transients and input activity, while these displays do not prove high-frequency jitter or instability.}\label{fig:activity}
\end{figure*}

Figure~\ref{fig:activity} asks whether the T32 position gain is accompanied by stronger angular or actuator activity. The upper panels use angular-rate error relative to the reference; the lower panels use changes in commanded motor speeds. These are different quantities even though both are expressed in rad/s. The plotted rate-error norm uses all three stored body components relative to their reference. Component/reference histories are retained in the plotting CSV. Both force application and withdrawal generate larger command excursions in these fixed H and F8 cases; the upper panels show larger angular-rate transients with compensation. The all-case summaries quantify input activity; the panels locate it in time. Neither substitutes for a spectral or physical vibration measurement. The supported conclusion remains increased angular-rate error and command activity alongside a position benefit.

\section{WN identities, computation, and experimental units}\label{sec:wn_details}

\begin{table*}[!t]
\caption{WN absolute case means. Bold minima refer only to the common 96-block comparison; 16-block structural rows are not ranked against it.}\label{tab:wn_all}\begin{center}\footnotesize\setlength{\tabcolsep}{3pt}\begin{tabular}{lrrrr}\toprule
Identity & Blocks & IAE (m\,$\cdot$\,s) & Peak (m) & End (mm)\\\midrule
Nominal NMPC & 96 & 0.107246 & 0.056001 & \textbf{0.265644}\\
Supervised cycle & 96 & 0.097677 & 0.053132 & 0.612527\\
Supervised stage & 96 & \textbf{0.094390} & \textbf{0.052195} & 0.452736\\
External RL Seed 1 & 96 & 0.097897 & 0.053162 & 0.638868\\

Internal cycle Seed 1 & 96 & 0.106968 & 0.053601 & 3.246368\\

Internal stage Seed 1 & 96 & 0.103561 & 0.052683 & 3.061903\\

Angular mask Seed 1 & 16 & 0.109491 & 0.055049 & 3.325548\\

Horizon decay Seed 1 & 16 & 0.106184 & 0.054621 & 2.530395\\

Retrained Seed 1 & 96 & 0.121256 & 0.057414 & 4.036391\\

PID & 96 & 0.269645 & 0.120826 & 3.166781\\\bottomrule\end{tabular}\end{center}
\end{table*}

\noindent\begin{minipage}{\columnwidth}
\captionof{table}{Pooled callback measurements, 88000 calls per row; deadlines include all samples. Bold marks observed column minima, not a causal timing advantage.}\label{tab:timing}\begin{center}\footnotesize\setlength{\tabcolsep}{3pt}\begin{tabular}{lrrrr}\toprule
Identity & P95 ms & $\ge20$ & $\ge50$ & Max ms\\\midrule
Nominal NMPC & \textbf{2.505} & 43 & 14 & 206.035\\
Supervised cycle & 2.733 & \textbf{10} & 6 & 172.190\\
Supervised stage & 3.338 & 11 & 3 & 305.087\\
Internal cycle Seed 1 & 3.143 & 12 & 4 & \textbf{86.467}\\
Internal stage Seed 1 & 3.804 & 14 & \textbf{2} & 128.330\\\bottomrule\end{tabular}\end{center}
\end{minipage}

\noindent\begin{minipage}{\columnwidth}
\captionof{table}{Available per-case P95 callback summaries for Seed 1 original policies; 220 cases per row. The mean and range of case P95 values are not pooled-callback P95 or maximum callback latency. External and internal recorded boundaries are not assumed identical.}\label{tab:seed_timing}
\centering\footnotesize\setlength{\tabcolsep}{3pt}
\begin{tabular}{lrrr}\toprule
Identity & Mean (ms) & Min (ms) & Max (ms)\\\midrule
External Seed 1 & 3.055 & 2.890 & 3.506\\

Internal cycle Seed 1 & 3.128 & 2.830 & 3.882\\

Internal stage Seed 1 & 3.774 & 3.369 & 4.706\\

\bottomrule\end{tabular}\end{minipage}

The grid case, WN environment block, and policy-training seed are different units. Time nodes and reused baselines are not extra replications. Fixed-grid contrasts and four WN realizations per condition are descriptive within the frozen bank. We do not conduct a pooled IID significance test or treat reused analysis pairs as independent trials. Pooled callback quantiles use recorded calls; average case P95 values are a different statistic. Separate acquisitions and unmatched T32 boundaries preclude a causal cross-batch overhead ranking. No latency is inserted into these synchronous simulated dynamics.

The archived manifests for both acquisitions underlying Figure~\ref{fig:fig08_callback_cdf} record an Intel Core i7-13650HX under WSL2 Linux 6.6.87.2, with OMP, MKL, and OpenBLAS thread environment variables each set to 1. CPU affinity covered logical CPUs 0--19; execution was not pinned to a single CPU. Python was 3.12.3, PyTorch 2.6.0+cpu, CasADi 3.7.2, and acados\_template 0.5.1. These records do not establish isolated execution or equal system load across batches. The 50 ms threshold is a separate project screening bound, not the 20 ms sampling deadline or a demonstrated physical stability limit.

\section{Setting rationale and evidence scope}\label{sec:settings}

Every exact value is classified as a literature/implementation principle (L), physical/numerical derivation (D), project evidence (V), or controlled-budget choice (P). These classes explain reproducibility without inventing optimality. Existing values are not retrospectively declared prespecified from literature. The accompanying setting register retains exact source/configuration paths and hashes.

\begin{table}[!t]
\caption{Setting classes and support boundaries; details remain in the hashed setting register.}\label{tab:setting_basis}\begin{center}\footnotesize\setlength{\tabcolsep}{3pt}\begin{tabular}{p{.18\columnwidth}p{.10\columnwidth}p{.59\columnwidth}}\toprule
Setting & Class & Basis\\\midrule
Plant/model & L/D/V & Iris source assembly; aerodynamic modeling \cite{r17,r18,r29,r30}.\\
Numerics & P/V & 50 Hz control, 200 Hz output, 20 nodes; fixed checked tolerances, not optimality.\\
NMPC & L/P & RTI/HPIPM principles \cite{r25,r26,r27}; common cost scales and limits.\\
Supervised fit & L/P/V & Residual principle \cite{r3,r7}; 10--32--6 and fitting caps are project choices.\\
Original SAC & L/P & Default LR, discount, target update \cite{r38}; other sizes use equal finite budget.\\
Action/\allowbreak reward & D/P/V & External 2\%; internal training quantile bounds; explicit 11-cost objective.\\
Seeds/\allowbreak load bank & P & Seed 1 (6201) for displayed policies; fixed coverage; finite-run limits \cite{r37,r40}.\\
Decay/\allowbreak filter/\allowbreak gain & D/P/V & Time constants derived; limited gain development; observer principle \cite{r41}.\\

\bottomrule\end{tabular}\end{center}
\end{table}

Physical settings use assembled inertia diagonal $(0.0300723,0.0306519,0.0576002)$ kg m$^2$ with off-diagonal entries retained; rotor thrust coefficient $5.84\times10^{-6}$ N/(rad/s)$^2$; motor bounds 0--1100 rad/s; acceleration/deceleration lag 0.0125/\allowbreak 0.025 s. The source vehicle has a 1.5 kg base and 1.55 kg assembled total; parameter assembly is not an independent identification \cite{r29}. Plant RK45 uses relative tolerance $10^{-12}$, 17-component absolute tolerance $10^{-15}$ and maximum internal step 625 $\mu$s. Prediction has eight 2.5 ms RK4 substeps per 20 ms interval. These common checked choices isolate method changes; no broad sensitivity study proves them optimal.

Original actor/critic networks use two 64-unit rectified linear unit (ReLU) layers. Original SAC uses Stable-Baselines3 2.7.0, PyTorch 2.6.0 CPU, learning rate 0.0003, discount 0.99, target update 0.005, batch 32, buffer 1600, warm-up 32 and 6400 interactions. Its automatic temperature starts at 0.1, with target entropy $-4$ external/$-6$ internal. Eligible interactions each receive one gradient update. The buffer holds four complete episodes; warm-up is 0.64 s. Internal scales are $(0.572,0.546,0.053,0.189,0.068,0.030)$, obtained by case-max componentwise training P95, upward rounding and floor 0.001. Original/default hyperparameters are not validated as sufficient.

Additional settings have explicit comparison roles. The 0.4 s horizon provides fixed finite preview at 50 Hz; it is held common to avoid attributing a larger optimization problem to the residual method. The 200 Hz output resolves each 20 ms control interval at four stored nodes, and the checked RK45/RK4 rules keep plant and predictor numerics distinct. No horizon or tolerance optimality is asserted. The cost scales nondimensionalize state blocks; its control, increment and terminal weights fix relative priorities across methods (Supplementary Section~\ref{sec:rl_details}).

The H/L/C/F8 references contrast static and moving mismatch with a common start, duration and yaw. Axis and oblique forces test directional dependence; nine positive strengths span the frozen finite bank. The 2--3 s pulse leaves two seconds before loading and five seconds after removal, separating pre-force, forcing and recovery errors. This defines the experiment, not an atmosphere or a complete angular envelope. The random-force harmonic band and tapered window supply a controlled time-varying counterpart at prescribed RMS loading; the four realizations give within-bank variation. The network receives none of these force histories.

For policy benchmarks, equal architecture and interaction budgets isolate the deployed interfaces. The buffer holds four complete episodes, the batch size bounds each update, and the 32-transition warm-up initializes replay; none proves sufficient training. The external 2\% hover-speed bound limits post-solve intervention. Internal bounds come only from the training residual quantiles, not test forces. Reward weights express the stated tracking/regularization priorities and failure penalties; they are not derived from the final IAE ranking. The 5 m abort and eight-second truncation bound runs, rather than certify physical safety. Seed identities fix reproducibility, not favorable selection \cite{r37,r38,r40}.

V1 and V2 isolate angular-output use and horizon persistence at unchanged actor weights. The observer filter is a finite development comparison, with its time constant derived in Supplementary Section~\ref{sec:fixed_confirmation}; gains are selected on D24 before T32. T32's force bands, timing ranges and strata test a selected compensator on newly generated conditions. Its 3\% gain, peak/tail tolerances and nonworse count are frozen engineering acceptance criteria, not literature-established universal thresholds. Metric and recovery definitions separate accumulated, extreme and late error; activity checks do not assert a frequency-domain vibration result. CFD settings and their source-versus-budget distinctions are specified separately in Table~\ref{tab:cfd_settings}.

Invalid infrastructure attempts, interrupted training, obsolete GN models and replaced actors are excluded by identity; valid unfavorable outcomes remain. Early fixed matrices are different protocols and remain in the engineering archive. Source hashes, weights, plot data and MATLAB sources are retained locally; public deposition is not complete.

\section{Absolute transient and recovery summaries}\label{sec:absolute_windows}
The 2--4 s peak includes one second after pulse removal; the 7--8 s mean measures persistent late error. They are reference errors, not differences between controllers or substitutes for full-window IAE. The external-policy rows use Seed 1 (6201). Means weight each of the 55 prescribed cases equally (one at zero force and six per positive intensity). The complete intensity curves remain in the engineering plotting package.

\noindent\begin{minipage}{\columnwidth}
\captionof{table}{Absolute transient and late errors (m). External policies closely match the cyclewise peak but increase the late error on these task means. No settling-time ranking is inferred.}\label{tab:absolute_windows}
\centering\small\setlength{\tabcolsep}{3pt}\begin{tabular}{llrr}\toprule
Task & Method & Peak [2,4] s & Mean [7,8] s\\\midrule
H & Nominal & 0.177960 & 0.000005\\
H & Cycle & 0.185253 & 0.000032\\
H & Stage & 0.180168 & 0.000019\\
H & External Seed 1 & 0.185208 & 0.000126\\

L & Nominal & 0.176318 & 0.000034\\
L & Cycle & 0.184470 & 0.000038\\
L & Stage & 0.179413 & 0.000020\\
L & External Seed 1 & 0.184396 & 0.000143\\

\bottomrule\end{tabular}\end{minipage}

A small transient gain can coexist with larger full-window IAE because late error accumulates. This is an arithmetic explanation supported by the time histories, not a mechanism that uniquely attributes the error to actor bias. The archived 0.05 m/0.5 s settling diagnostic is not used to assert a recovery-time ranking.

\clearpage
\begin{figure*}[p]
\begin{minipage}{\textwidth}\section{Remaining tasks and policy curves}\label{sec:other_tasks}\end{minipage}
\centering
\begin{subfigure}[t]{.48\textwidth}\centering
\includegraphics[width=\linewidth,height=38mm,keepaspectratio]{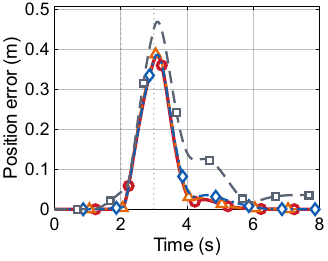}
\caption{C: pulse, $\lambda=0.24$, horizontal $+45^\circ$}\label{fig:other_C_pulse}
\end{subfigure}\hfill
\begin{subfigure}[t]{.48\textwidth}\centering
\includegraphics[width=\linewidth,height=38mm,keepaspectratio]{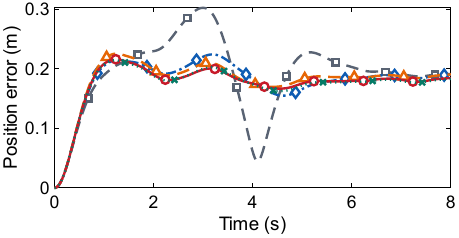}
\caption{C: continuous CFD inflow}\label{fig:cfd_other}
\end{subfigure}
\basiclegend
\par{\small In (b), blue is the same-graph zero-residual control; green dotted $\times$ is internal stagewise RL.}
\caption{C histories complete the main H/L/F8 comparison. Panel (a) uses four basic controllers under the pulse protocol; panel (b) uses five frozen controllers in the CFD field. Supervised recovery gains after pulse removal differ from the persistent offsets under continuous inflow. These are separate protocols, not paired realizations of one load.}\label{fig:other_tasks}
\par\smallskip\noindent Figure~\ref{fig:other_tasks}\,\subref{fig:other_C_pulse} shows lower late pulse error with supervision, without eliminating the forcing transient. Panel~\subref{fig:cfd_other} instead retains a sustained offset: stagewise supervision reduces C IAE relative to zero residual, whereas cyclewise does not. Internal stagewise RL further reduces position IAE but raises angular-rate RMSE relative to stagewise supervision. Tables~\ref{tab:all_grid}, \ref{tab:cfd_position}, \ref{tab:cfd_secondary} retain the respective full comparisons.
\par\medskip\nextfloat

\begin{subfigure}[t]{.48\textwidth}
  \centering
  \includegraphics[width=\linewidth]{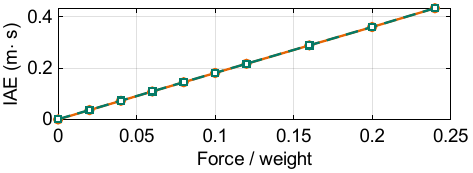}
  \caption{H: external RL}\label{fig:si_intensity_1_H}
\end{subfigure}\hfill\begin{subfigure}[t]{.48\textwidth}
  \centering
  \includegraphics[width=\linewidth]{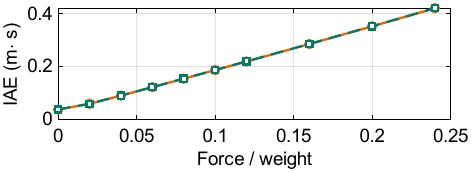}
  \caption{F8: external RL}\label{fig:si_intensity_1_F8}
\end{subfigure}
\par\smallskip
\begin{subfigure}[t]{.48\textwidth}
  \centering
  \includegraphics[width=\linewidth]{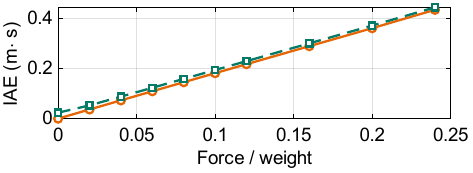}
  \caption{H: internal cyclewise RL}\label{fig:si_intensity_2_H}
\end{subfigure}\hfill\begin{subfigure}[t]{.48\textwidth}
  \centering
  \includegraphics[width=\linewidth]{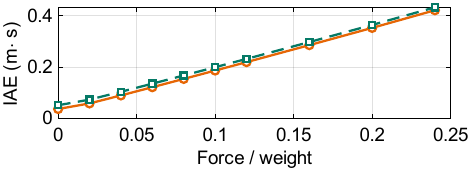}
  \caption{F8: internal cyclewise RL}\label{fig:si_intensity_2_F8}
\end{subfigure}
\par\smallskip
\begin{subfigure}[t]{.48\textwidth}
  \centering
  \includegraphics[width=\linewidth]{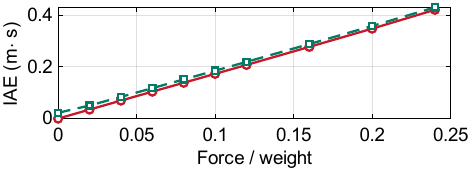}
  \caption{H: internal stagewise RL}\label{fig:si_intensity_3_H}
\end{subfigure}\hfill\begin{subfigure}[t]{.48\textwidth}
  \centering
  \includegraphics[width=\linewidth]{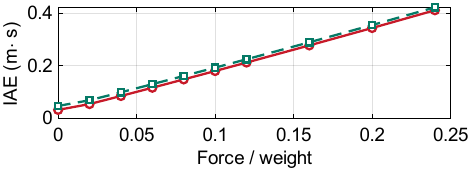}
  \caption{F8: internal stagewise RL}\label{fig:si_intensity_3_F8}
\end{subfigure}
\par\smallskip
{\small\curvekey{cyclecolor}{solid}{$\circ$}{Cyclewise baseline}\curvekey{stagecolor}{solid}{$\circ$}{Stagewise baseline}\curvekey{rlcolor}{dashed}{$\square$}{Matched RL}}
\caption{Absolute intensity curves retain Seed 1 (6201) for all three interfaces. Positive points average six directions; zero-force points are single cases. External effects are small on this scale; internal losses depend on task. The complete-grid tables, not these selected task displays, determine the aggregate conclusion.}\label{fig:rl_intensity_full}
\par\smallskip\noindent Figure~\ref{fig:rl_intensity_full} shows that increasing force strength raises absolute IAE in both methods of each matched pair. External RL nearly overlaps cyclewise supervision; the internal-policy curves retain larger errors in H and F8. Thus reducing the plotted range to selected intensities would not establish an RL benefit. These taskwise direction means support the full-grid result in Table~\ref{tab:all_grid}, without implying that every individual direction has the same ordering.
\end{figure*}

\begin{figure*}[p]\centering
\begin{subfigure}[t]{.32\textwidth}\centering
\includegraphics[width=\linewidth]{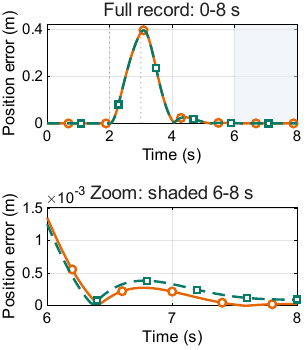}
\caption{H: external RL}\label{fig:rl_1_H}\end{subfigure}\hfill\begin{subfigure}[t]{.32\textwidth}\centering
\includegraphics[width=\linewidth]{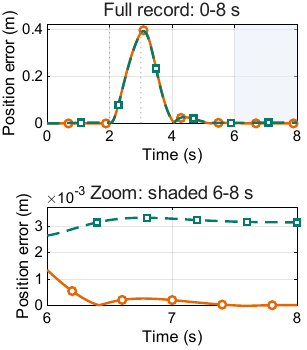}
\caption{H: internal cyclewise RL}\label{fig:rl_2_H}\end{subfigure}\hfill\begin{subfigure}[t]{.32\textwidth}\centering
\includegraphics[width=\linewidth]{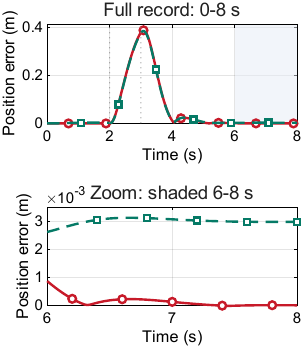}
\caption{H: internal stagewise RL}\label{fig:rl_3_H}\end{subfigure}
\par\medskip
\begin{subfigure}[t]{.32\textwidth}\centering
\includegraphics[width=\linewidth]{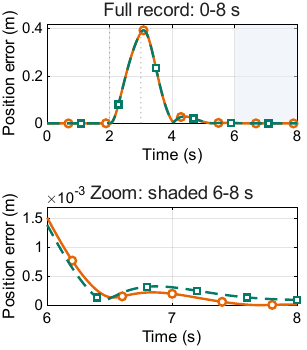}
\caption{L: external RL}\label{fig:rl_1_L}\end{subfigure}\hfill\begin{subfigure}[t]{.32\textwidth}\centering
\includegraphics[width=\linewidth]{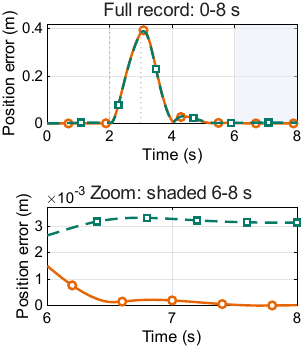}
\caption{L: internal cyclewise RL}\label{fig:rl_2_L}\end{subfigure}\hfill\begin{subfigure}[t]{.32\textwidth}\centering
\includegraphics[width=\linewidth]{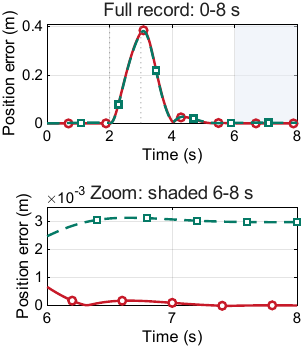}
\caption{L: internal stagewise RL}\label{fig:rl_3_L}\end{subfigure}
\rllegend
\caption{H and L complete the matched policy histories at $\lambda=0.24$, horizontal $+45^\circ$, for Seed 1 (6201). Each panel shows the full 0--8 s record and its shaded 6--8 s interval, both in m. No samples are shifted or smoothed. External RL stays close to its supervised baseline; the internal policies leave larger late offsets in both tasks.}\label{fig:rl_other_histories}
\par\smallskip\noindent Figure~\ref{fig:rl_other_histories} tests whether the late behavior in the main F8 plot also appears in H and L. Panels (a), (d) show small external-policy differences despite nearly coincident peaks. Panels (b), (c), (e), (f) show persistent internal-policy tail errors after the main disturbance transient has subsided. Hence the F8 tail observation is not confined to that trajectory, although these six histories remain selected cases; Table~\ref{tab:all_grid} supplies the full-grid comparison. Lower plots are magnified windows of the same position-error norm, not a different error metric.
\end{figure*}

\begin{figure*}[p]\centering\setlength{\parskip}{0pt}\captionsetup{skip=8pt}
\begin{minipage}{\textwidth}
\section{Synthetic CFD field and matched records}\label{sec:cfd_details}
\end{minipage}
\begin{minipage}{\textwidth}
\captionof{table}{CFD settings: exact case values and bounded reasons. P denotes a project choice, D a numerical definition, and L the cited implementation.}\label{tab:cfd_settings}
\centering\small\setlength{\tabcolsep}{3pt}
\begin{tabular}{p{.22\textwidth}p{.71\textwidth}}\toprule
Setting & Value, purpose and basis\\\midrule
Domain / mesh & Periodic 8 m cube; $64^3$ cells, 0.125 m spacing. P: bounded field-storage and solver budget; no mesh-convergence claim.\\
Mean / fluctuation & $(3,0,0)$ / 0.75 m/s vector RMS initially. P: prescribed persistent cross-flow and fluctuation amplitude, not atmospheric calibration.\\
Initialization & Seed 73001; analytic spectrum with $k_0=4$ rad/m. P: reproducible spatial variation, not a random population sample.\\
Viscosity & $1.5\times10^{-5}$ m$^2$/s. P: prescribed air-like kinematic viscosity; no temperature calibration.\\
WALE & $C_w=0.325$, $C_k=0.094$, cube-root-cell-volume filter width. L: v2512 defaults \cite{r47}. Inherited $C_e=1.048$ is recorded, not an independently tuned WALE mechanism.\\
Time / output & Fixed 0.005 s, 1600 steps; 0.02 s output, 401 fields. P/D: eight-second comparison; maximum logged Courant number about 0.263. Short half-step probes do not establish full-run convergence.\\
Queries & Periodic trilinear space / linear physical time; no temporal wrap. D: regular cell-centred bank, rotor queries every 5 ms.\\
Controller & Same 20 ms controller, frozen $g_\phi$, zero residual comparator on same stagewise graph. P: compare five frozen methods within this protocol; no predictor or actor wind input.\\\bottomrule
\end{tabular}\end{minipage}
\par

\begin{minipage}[t]{.48\textwidth}\captionof{table}{All 20 CFD position records. Bold is the taskwise minimum among the five methods, not a cross-protocol ranking. IAE is in m\,$\cdot$\,s; remaining columns in m.}\label{tab:cfd_position}\centering\small\setlength{\tabcolsep}{4pt}\begin{tabular}{llrrrr}\toprule Task & Method & IAE & RMSE & Peak & End\\\midrule
H & PID & 1.442387 & 0.184422 & 0.216801 & 0.191019\\
H & Zero & 1.430646 & 0.182750 & 0.211330 & 0.190863\\
H & Cycle & 1.442156 & 0.184307 & 0.222114 & 0.191321\\
H & Stage & 1.395343 & 0.178235 & 0.212760 & 0.185260\\
H & RL S1 & \textbf{1.383999} & \textbf{0.176765} & \textbf{0.210999} & \textbf{0.183659}\\
L & PID & 1.399181 & 0.178987 & 0.211222 & 0.190482\\
L & Zero & 1.395860 & 0.178383 & 0.211283 & 0.189486\\
L & Cycle & 1.429108 & 0.182666 & 0.222151 & 0.189675\\
L & Stage & 1.383166 & 0.176707 & 0.212832 & 0.183718\\
L & RL S1 & \textbf{1.368012} & \textbf{0.174760} & \textbf{0.211058} & \textbf{0.182254}\\
C & PID & 1.534373 & 0.202267 & 0.302553 & 0.191380\\
C & Zero & 1.442937 & 0.184899 & 0.224282 & 0.190305\\
C & Cycle & 1.452657 & 0.185832 & 0.225652 & 0.190683\\
C & Stage & 1.405305 & 0.179632 & 0.216229 & 0.184663\\
C & RL S1 & \textbf{1.395709} & \textbf{0.178437} & \textbf{0.214461} & \textbf{0.183138}\\
F8 & PID & 1.524495 & 0.202703 & 0.383735 & 0.191930\\
F8 & Zero & 1.433735 & 0.183467 & 0.218057 & 0.190785\\
F8 & Cycle & 1.448715 & 0.185252 & 0.222255 & 0.191228\\
F8 & Stage & 1.404637 & 0.179520 & 0.212924 & 0.185183\\
F8 & RL S1 & \textbf{1.394309} & \textbf{0.178226} & \textbf{0.211199} & \textbf{0.183597}\\
\bottomrule\end{tabular}\end{minipage}\hfill
\begin{minipage}[t]{.48\textwidth}\captionof{table}{All 20 CFD secondary RMSE records. Bold marks the taskwise minimum; RL S1 means the retained internal stagewise actor, Seed 1.}\label{tab:cfd_secondary}\centering\small\setlength{\tabcolsep}{4pt}\begin{tabular}{llrrr}\toprule Task & Method & Velocity (m/s) & Attitude (rad) & Rate (rad/s)\\\midrule
H & PID & \textbf{0.076284} & \textbf{0.112631} & 0.098406\\
H & Zero & 0.082519 & 0.113241 & 0.090664\\
H & Cycle & 0.086999 & 0.113414 & \textbf{0.088828}\\
H & Stage & 0.084243 & 0.113379 & 0.090490\\
H & RL S1 & 0.083922 & 0.113388 & 0.090794\\
L & PID & \textbf{0.077422} & \textbf{0.110318} & 0.099189\\
L & Zero & 0.082853 & 0.110825 & 0.090854\\
L & Cycle & 0.087190 & 0.110998 & \textbf{0.088985}\\
L & Stage & 0.084407 & 0.110960 & 0.090678\\
L & RL S1 & 0.084093 & 0.110963 & 0.090939\\
C & PID & 0.192835 & 0.126608 & 0.195537\\
C & Zero & 0.091939 & 0.114838 & 0.098126\\
C & Cycle & 0.092812 & 0.115024 & \textbf{0.097305}\\
C & Stage & \textbf{0.089066} & \textbf{0.114818} & 0.097331\\
C & RL S1 & 0.089150 & 0.114868 & 0.097757\\
F8 & PID & 0.280043 & 0.165102 & 0.564603\\
F8 & Zero & 0.093504 & 0.114097 & 0.103961\\
F8 & Cycle & 0.093395 & 0.114259 & 0.103977\\
F8 & Stage & \textbf{0.090186} & \textbf{0.114033} & \textbf{0.101070}\\
F8 & RL S1 & 0.090250 & 0.114040 & 0.101595\\
\bottomrule\end{tabular}\end{minipage}

\par

\begin{minipage}[t]{.48\textwidth}
OpenFOAM v2512 pimpleFoam evolves a periodic incompressible field with the WALE closure \cite{r46,r47}. Initialization uses a prescribed radial spectral envelope proportional to $k^4\exp[-2(k/k_0)^2]$, with angular wavenumber $k$ in rad/m and $k_0=4$ rad/m. A discrete divergence-free projection precedes amplitude normalization and mean-flow addition. Although the initialization utility has a historical CBC name, this case supplies an analytic spectrum, not wind-tunnel measurements. Its vector fluctuation RMS is $[\langle\|\mathbf w-\langle\mathbf w\rangle\|_2^2\rangle]^{1/2}$, where brackets denote a spatial cell average; this is not a single-component turbulence-intensity statistic.

The mean flow remains about 3 m/s, while vector fluctuation RMS decays from 0.750 to 0.310 m/s. No sustained fluctuation forcing is applied. Continuity and Courant checks support execution integrity, not atmospheric fidelity. The archived field views were produced in ParaView 6.1 \cite{r48}. Figure~\ref{fig:cfd} is redrawn in MATLAB from the same bank, evaluating speed after vector interpolation to $z=1$ m.
\end{minipage}\hfill
\begin{minipage}[t]{.48\textwidth}
Physical-plane LES views follow \cite{r50}; the sequential color map follows \cite{r49}. Both views use the same 4096 samples. The plane is regenerated from the archived velocity bank, without a screenshot or smoothing. Its labels, axes and arrows are vector objects, and all $64^2$ cloud cells are retained. It keeps the 1--5.5 m/s scale; the speed surface uses 2--4 m/s. Surface height denotes speed, not elevation. Black arrows show local horizontal velocity, including transverse fluctuations. The Reference box bounds the four horizontal reference paths with 0.05 m padding; it is not a vehicle footprint. Periodic cell-centre coordinates are preserved.

All 20 trajectories complete without fallback. Independent audits and 40 paired records do not add performance samples. Final verification binds the field, inputs, weights and logs to their hashes. Failed assemblies and interruptions are excluded from performance counts but remain archived.
\end{minipage}
\end{figure*}
\clearpage

\end{document}